\documentclass[aps,prc,superscriptaddress,twocolumn,nofootinbib,amsmath,amssymb,showpacs,showkeys]{revtex4-1}
\usepackage{pifont}
\usepackage{graphicx}
\usepackage{color}
\usepackage{amsmath}
\usepackage{amssymb}
\usepackage[caption=false]{subfig}
\usepackage{multirow}
\usepackage{longtable}

\newcommand{\proton}{\ensuremath{p}}
\newcommand{\neutron}{\ensuremath{n}}
\newcommand{\pip}{\ensuremath{\pi^{+}}}
\newcommand{\pim}{\ensuremath{\pi^{-}}}
\newcommand{\kp}{\ensuremath{K^{+}}}

\newcommand{\pizero}{\ensuremath{\pi^{0}}}
\newcommand{\SigmaPlus}{\ensuremath{\Sigma^{+}}}
\newcommand{\SigmaZero}{\ensuremath{\Sigma^{0}}}
\newcommand{\SigmaMinus}{\ensuremath{\Sigma^{-}}}
\newcommand{\LambdaOne}{\ensuremath{\Lambda(1405)}}
\newcommand{\LambdaTwo}{\ensuremath{\Lambda(1520)}}

\newcommand{\mevcc}{\ensuremath{\mathrm{MeV}/c^{2}}} 
\newcommand{\gevcc}{\ensuremath{\mathrm{GeV}/c^{2}}} 

\newcommand{\deltaTOF}{\ensuremath{\Delta \mathrm{TOF}}}

\newcommand{\costhetakp}{\ensuremath{\cos \theta_{\kp}^{\mathrm{c.m.}}}}

\newcommand{\sigmatot}{\ensuremath{\sigma_{\mathrm{tot}}}}

\newcommand{\etal}{\textit{et al.}}

\begin{document}



\preprint{CMU/101-2013}

%
%
%
%
%
%
%
%
%
%

\newcommand*{\CMU}{Carnegie Mellon University, Pittsburgh, Pennsylvania 15213}
\affiliation{\CMU}
\newcommand*{\ANL}{Argonne National Laboratory, Argonne, Illinois 60439}
\affiliation{\ANL}
\newcommand*{\ASU}{Arizona State University, Tempe, Arizona 85287-1504}
\affiliation{\ASU}
\newcommand*{\CSUDH}{California State University, Dominguez Hills, Carson, CA 90747}
\affiliation{\CSUDH}
\newcommand*{\CANISIUS}{Canisius College, Buffalo, NY}
\affiliation{\CANISIUS}
\newcommand*{\CUA}{Catholic University of America, Washington, D.C. 20064}
\affiliation{\CUA}
\newcommand*{\SACLAY}{CEA, Centre de Saclay, Irfu/Service de Physique Nucl\'eaire, 91191 Gif-sur-Yvette, France}
\affiliation{\SACLAY}
\newcommand*{\CNU}{Christopher Newport University, Newport News, Virginia 23606}
\affiliation{\CNU}
\newcommand*{\UCONN}{University of Connecticut, Storrs, Connecticut 06269}
\affiliation{\UCONN}
\newcommand*{\EDINBURGH}{Edinburgh University, Edinburgh EH9 3JZ, United Kingdom}
\affiliation{\EDINBURGH}
\newcommand*{\FU}{Fairfield University, Fairfield CT 06824}
\affiliation{\FU}
\newcommand*{\FIU}{Florida International University, Miami, Florida 33199}
\affiliation{\FIU}
\newcommand*{\FSU}{Florida State University, Tallahassee, Florida 32306}
\affiliation{\FSU}
\newcommand*{\Genova}{Universit$\grave{a}$ di Genova, 16146 Genova, Italy}
\affiliation{\Genova}
\newcommand*{\GWUI}{The George Washington University, Washington, DC 20052}
\affiliation{\GWUI}
\newcommand*{\ISU}{Idaho State University, Pocatello, Idaho 83209}
\affiliation{\ISU}
\newcommand*{\INFNFE}{INFN, Sezione di Ferrara, 44100 Ferrara, Italy}
\affiliation{\INFNFE}
\newcommand*{\INFNFR}{INFN, Laboratori Nazionali di Frascati, 00044 Frascati, Italy}
\affiliation{\INFNFR}
\newcommand*{\INFNGE}{INFN, Sezione di Genova, 16146 Genova, Italy}
\affiliation{\INFNGE}
\newcommand*{\INFNRO}{INFN, Sezione di Roma Tor Vergata, 00133 Rome, Italy}
\affiliation{\INFNRO}
\newcommand*{\ORSAY}{Institut de Physique Nucl\'eaire ORSAY, Orsay, France}
\affiliation{\ORSAY}
\newcommand*{\ITEP}{Institute of Theoretical and Experimental Physics, Moscow, 117259, Russia}
\affiliation{\ITEP}
\newcommand*{\JMU}{James Madison University, Harrisonburg, Virginia 22807}
\affiliation{\JMU}
\newcommand*{\KNU}{Kyungpook National University, Daegu 702-701, Republic of Korea}
\affiliation{\KNU}
\newcommand*{\LPSC}{LPSC, Universite Joseph Fourier, CNRS/IN2P3, INPG, Grenoble, France}
\affiliation{\LPSC}
\newcommand*{\UNH}{University of New Hampshire, Durham, New Hampshire 03824-3568}
\affiliation{\UNH}
\newcommand*{\NSU}{Norfolk State University, Norfolk, Virginia 23504}
\affiliation{\NSU}
\newcommand*{\OHIOU}{Ohio University, Athens, Ohio  45701}
\affiliation{\OHIOU}
\newcommand*{\ODU}{Old Dominion University, Norfolk, Virginia 23529}
\affiliation{\ODU}
\newcommand*{\RPI}{Rensselaer Polytechnic Institute, Troy, New York 12180-3590}
\affiliation{\RPI}
\newcommand*{\URICH}{University of Richmond, Richmond, Virginia 23173}
\affiliation{\URICH}
\newcommand*{\ROMAII}{Universita' di Roma Tor Vergata, 00133 Rome Italy}
\affiliation{\ROMAII}
\newcommand*{\MSU}{Skobeltsyn Nuclear Physics Institute, 119899 Moscow, Russia}
\affiliation{\MSU}
\newcommand*{\SCAROLINA}{University of South Carolina, Columbia, South Carolina 29208}
\affiliation{\SCAROLINA}
\newcommand*{\JLAB}{Thomas Jefferson National Accelerator Facility, Newport News, Virginia 23606}
\affiliation{\JLAB}
\newcommand*{\UNIONC}{Union College, Schenectady, NY 12308}
\affiliation{\UNIONC}
\newcommand*{\UTFSM}{Universidad T\'{e}cnica Federico Santa Mar\'{i}a, Casilla 110-V Valpara\'{i}so, Chile}
\affiliation{\UTFSM}
\newcommand*{\GLASGOW}{University of Glasgow, Glasgow G12 8QQ, United Kingdom}
\affiliation{\GLASGOW}
\newcommand*{\VIRGINIA}{University of Virginia, Charlottesville, Virginia 22901}
\affiliation{\VIRGINIA}
\newcommand*{\WM}{College of William and Mary, Williamsburg, Virginia 23187-8795}
\affiliation{\WM}
\newcommand*{\YEREVAN}{Yerevan Physics Institute, 375036 Yerevan, Armenia}
\affiliation{\YEREVAN}

\newcommand*{\NOWLANL}{Los Alamos National Laboratory, Los Alamos, NM 87544 USA}
\newcommand*{\NOWUCONN}{University of Connecticut, Storrs, Connecticut 06269}
\newcommand*{\NOWMSU}{Skobeltsyn Nuclear Physics Institute, 119899 Moscow, Russia}
\newcommand*{\NOWORSAY}{Institut de Physique Nucl\'eaire ORSAY, Orsay, France}
\newcommand*{\NOWINFNGE}{INFN, Sezione di Genova, 16146 Genova, Italy}
\newcommand*{\NOWROMAII}{Universita' di Roma Tor Vergata, 00133 Rome Italy}
\newcommand*{\NOWINDIANA}{Indiana University, Bloomington, Indiana 47405}
\newcommand*{\NOWSIENA}{Siena College, Loudonville, NY 12211}

\author {K.~Moriya} 
\altaffiliation[Current address:]{\NOWINDIANA}
\affiliation{\CMU}
\author {R.A.~Schumacher} 
\affiliation{\CMU}
\author {K.P. ~Adhikari} 
\affiliation{\ODU}
\author {D.~Adikaram} 
\affiliation{\ODU}
\author {M.~Aghasyan} 
\affiliation{\INFNFR}
\author {M.J.~Amaryan} 
\affiliation{\ODU}
\author {M.D.~Anderson} 
\affiliation{\GLASGOW}
\author {S. ~Anefalos~Pereira} 
\affiliation{\INFNFR}
\author {H.~Avakian} 
\affiliation{\JLAB}
\author {J.~Ball} 
\affiliation{\SACLAY}
% ...

\collaboration{The CLAS Collaboration}
\noaffiliation

%
%
%
%
%
%


\title{Differential Photoproduction Cross Sections of the 
$\mathbf{\Sigma^{0}(1385)}$, $\mathbf{\Lambda(1405)}$, and $\mathbf{\Lambda(1520)}$}





\date{\today}




\begin{abstract}
We report the exclusive photoproduction cross sections for the $\Sigma^{0}(1385)$,
\LambdaOne, and \LambdaTwo{} in the reactions $\gamma + p \to K^+ + Y^{\ast}$ 
using the CLAS detector for energies from near the respective
production thresholds up to a center-of-mass energy $W$ of $2.85$
GeV. The differential cross sections are integrated to give the total
exclusive cross sections for each hyperon.  Comparisons are made to
current theoretical models based on the effective Lagrangian approach
and fitted to previous data. The accuracy of these models is seen to
vary widely. The cross sections for the \LambdaOne{} region are
strikingly different for the $\Sigma^+\pi^-$, $\Sigma^0\pi^0$, and
$\Sigma^-\pi^+$ decay channels, indicating the effect of isospin
interference, especially at $W$ values close to the threshold.
\end{abstract}

\pacs{
      {13.30.Eg}    
      {13.40.-f}    
      {13.60.Rj}    
      {14.20.Gk}    
     } 



\maketitle

\section{Introduction}
\label{section:Introduction}

The \LambdaOne{} $(J^P=1/2^-)$, situated just below the $N \bar{K}$
threshold, has been an enigmatic state in the spectrum of strange
baryons for decades.  The differential photoproduction cross sections
should provide information needed to identify the dynamics that
play a significant role in the formation of the \LambdaOne, and lead
to a deeper understanding of any other structures that populate this
mass region. The \LambdaOne{} sits between two other well-known
hyperons, the $\Sigma^{0}(1385)$ ($J^P=3/2^+$) and the \LambdaTwo{} ($J^P =
3/2^-$).  Also for these states, photoproduction data have been scarce,
and comparison of the three hyperons was not practical until now.

The CLAS measurements reported here are the first to have been made
for all three excited hyperon species at the same time with the same
apparatus and within the same analysis.  A critical comparison of the
cross sections for the combined set of hyperons, including the ground
states, may yield further insight into their structures.  In
particular, since the \LambdaOne{} does not fit quantitatively into
quark model estimations of its
mass~\cite{Isgur-Karl_PRD18,Capstick:2000qj}, and since it has always
been thought to be strongly influenced by the nearby $N \bar{K}$ and
$\Sigma\pi$ thresholds~\cite{Dalitz_Tuan:PRL,Dalitz_Tuan:AnnPhys10},
one might expect differences in the mechanism of its production in
comparison to the more typical hyperons.

The outline of this paper is as follows.
Section~\ref{section:past_experiments} briefly describes several
previous hyperon photoproduction experiments, together with several
theoretical models that have been proposed.
Section~\ref{section:SetupAndAnalysis}
outlines the setup of the experiment and the steps taken toward cross
section extractions.
Subsection~\ref{subsection:setup:yields1385}
discusses the yield extraction for the
$\Sigma^{0}(1385)$ via the $\Lambda\pi$ decay mode
and Subsection~\ref{subsection:setup:yields1520_1405}
discusses the yields of the \LambdaOne{} and
\LambdaTwo{} via the three different $\Sigma\pi$ decay modes.   
Subsection~\ref{subsection:setup:acceptancenormalization} then discusses
the acceptance calculations computed using a Monte Carlo (MC) method, and
the photon beam flux normalization.
The systematic uncertainties of the cross section results will be
discussed in Subsection~\ref{subsection:setup:systematics}. 
Section~\ref{section:results} shows the results for each hyperon
individually, and then compares the hyperons to each other.
We recapitulate the main results in 
Sec.~\ref{section:Conclusions}.

\section{Past experiments and current theory}
\label{section:past_experiments}


For photoproduction of the $\Sigma^0(1385)$ on the proton there are
total cross
section measurements from bubble chamber work by the ABBHHM
group~\cite{Erbe:1970cq, Erbe:1967} and by the CEA
group~\cite{Crouch:1967zz}.  Preliminary results from CLAS have been
presented~\cite{Guo-1385}, but the present analysis is independent of
that study, albeit using the same raw data set.  In all these
measurements the hyperon was
reconstructed through its dominant decay to $\Lambda\pi^0$.
Alternatively, the LEPS Collaboration measured $\gamma n \to K^+
\Sigma^-(1385)$ using a deuteron target and detected only the \kp{}
and the $\pi^-$ from the hyperon decay to $\Lambda \pi^-$
~\cite{Hicks:2008yn}.  
The LEPS result showed a flat angular distribution for
$\costhetakp >0.6$ and an energy dependence that rose from
threshold to a wide peak near $W=1.8$ GeV.  That result was compared
with an effective Lagrangian model of Oh~\etal~\cite{Oh:2007jd} in which
the dominant contribution came from $t$-channel $K$ exchange and
little from $K^*$. Agreement was at best fair, both for the cross
section and for the beam asymmetry.  
One may expect comparable cross sections in the
reaction $\gamma p \to K^+ \Sigma^0(1385) $, as will be
discussed.  
We will compare our results for
this channel with the same model calculation in
Sec.~\ref{subsection:results:results1385}.


For photoproduction of the \LambdaTwo{} there are previous
experimental data from Boyarski \etal{} (SLAC)~\cite{Boyarski:1970yc}
at $E_\gamma = 11$~GeV and Barber \etal{} (LAMP2, Daresbury)
~\cite{Barber} at $E_\gamma=2.8-4.8$~GeV.  In more recent times, the
LEPS Collaboration has looked at photoproduction of this hyperon in
the energy region $1.9 < E_\gamma < 2.8 $ GeV using a
forward-angle spectrometer~\cite{Muramatsu:2009zp,Kohri:2009xe}.  They
showed that the cross section is forward peaked, and that this
behavior is more consistent with a model dominated by a
gauge-invariance-preserving contact term~\cite{Nam:2005uq,Nam:2010au},
and less consistent with models dominated by $t-$channel vector meson $K^*$
exchange~\cite{Titov:2006hz, Sibirtsev:2005ns}.  However, the
$K^{\ast}$ exchange models are
more consistent with the results at higher photon energies
(Ref.~\cite{Barber}). The beam asymmetry for the \LambdaTwo{} was found
to be small, much smaller than for the ground state $\Lambda$,
supporting the contact-term model that found $K^*$ exchange is not
important in the threshold region.  Furthermore, it was found that the
energy dependence of the forward-angle cross section rises from
threshold to a maximum near $W=2.15$ GeV, followed by a decline.  It
was suggested that this could be an effect of an $N^*$ intermediate
resonance at 2.11 GeV~\cite{Nam:2010au}.  The results in
the present paper cover a broader kinematic range
than previous data, and will be
compared to the approach in Ref.~\cite{Nam:2010au} and also
Ref.~\cite{JunHe} in Sec.~\ref{section:results}.  

Additional \LambdaTwo{} photoproduction data close to threshold were
published recently by SAPHIR~\cite{Wieland:2011zz}, to be discussed
later.  Pioneering measurements at Cornell~\cite{Mistry} and
CEA~\cite{Blanpied:1965zz} will not be discussed.  There is also a
\LambdaTwo{} electroproduction result from CLAS~\cite{Barrow:2001ds}
at $Q^2 > 0.9$ GeV/c$^2$ that we will not discuss here.


For photoproduction of the \LambdaOne{} there was very little
information up to now.  The LEPS Collaboration produced the only
significant measurement so far~\cite{Niiyama}.  They estimated that the
differential cross section for $0.8 < \costhetakp < 1.0$
and $1.5 < E_\gamma < 2.0 $ GeV is about 0.4~$\mu$b (no error estimate
given, but we suppose it to be $\sim 50\%$).   They also reported a
steep decrease in \LambdaOne{} production versus $\Sigma^{0}(1385)$
production in the higher energy range $2.0 < E_\gamma <
2.4$~GeV.  They speculated that this might be a hint of strong
dynamical differences in the production mechanisms.  We will consider
these findings in Sec.~\ref{subsection:results:results1405}.


Considering recent theoretical approaches, the photoproduction cross
section of the $\Sigma^{0}(1385)$ has been studied in an effective
Lagrangian model of Oh~\etal~\cite{Oh:2007jd}.  They pointed out that
since the cross section for this excited hyperon is comparable in
size to those of the ground state $\Lambda$ and $\Sigma^{0}$, it also
may serve as a hunting
ground for high-mass non-strange resonances that may
couple to it. 
The calculation was evaluated at tree-level, with
single-channel Born and resonance contributions using empirically
obtained couplings.  A set of four high mass $\Delta$ and $N^*$
resonances was found to contribute to the total cross section for the
reaction in the threshold region.  However, only a preliminary CLAS
total cross section result~\cite{Guo-1385} was available to fit, and
the resonances played a secondary role in matching the data.  The
contact interaction used to preserve gauge invariance was dominant.
Thus, the present paper that shows differential cross sections as well
as the integrated total cross section should help clarify the
theoretical modeling of this reaction, including any resonant
content.


The photoproduction cross section of the \LambdaTwo{} has been studied
theoretically by Nam \etal~\cite{Nam:2005uq}.  In an effective
Lagrangian approach, using Born terms and a
Rarita-Schwinger formalism for inclusion of the spin-3/2 \LambdaTwo,
they tested various model assumptions against scant previous data
~\cite{Barber} for the total cross section and the $t$-dependence at
$E_\gamma = 3.8$~GeV.  They found that the total cross
section near threshold was mainly determined by the contact interaction
included in order to preserve gauge invariance in the presence of
hadronic form factors.  They found only minor
sensitivity, for photon energies below 3 GeV, to the anomalous
magnetic moment of the hyperon, $\kappa_{\Lambda^\ast}$, and the
coupling constant $g_{K^\ast N \Lambda^\ast}$.  No $N^*$ ($s$-channel)
resonances were found to be necessary to qualitatively reproduce the
cross sections.  In a subsequent paper~\cite{Nam:2010au} the effect of
Reggeizing the $t-$channel exchanges was studied. At CLAS energies
(i.e. for $E_\gamma < 3.7$ GeV) this effect was found to be negligible.

In another effective Lagrangian approach model for the \LambdaTwo{}
cross section, He and Chen studied systematically the inclusion or
exclusion of several higher-mass Constituent Quark Model (CQM) nucleon
resonances~\cite{JunHe}.  Their approach was to fit the differential
cross sections from LEPS and also the higher energy $t$-dependence
from SLAC.  They concluded that the contact term is the dominant
contribution to the cross section at all energies.  Also, both $K$ and
$K^{*}$ exchanges play a significant role, though the $K^{*}$ only at
the higher SLAC energies. They found significant resonance
contribution only from the two-star $N(2080) D_{13}$, and a possible
hint that a CQM $D_{15}$ resonance at nearly the same mass could be
needed.  We note in passing that the $N(2080)D_{13}$ is a state that
is important in photoproduction of the ground-state
$\Lambda(1116)$~\cite{Schumacher:2010qx}.  (Also note that the most
recent Particle Data Group (PDG)~\cite{Beringer:1900zz} evaluation of
the $N^* 3/2^- D_{13} $ partial wave splits this state into two: one
at 1875 MeV and one at 2120 MeV.)  Their affirmative conclusion about
the need for these resonances is in disagreement with the previous
work by Nam \etal{} cited in Ref.~\cite{Kohri:2009xe}, and their
quantitative agreement with the $\LambdaTwo$ differential cross
section and with the beam asymmetry were not good.  We can expect that
the more complete angular distributions presented below will be useful
in refining these models.


The photoproduction cross section of the \LambdaOne{} has been studied
theoretically by Nam \etal~\cite{Nam:2008jy} within the framework of
the same model as discussed above for the \LambdaTwo.
Within their effective Lagrangian approach there is a mass cutoff in
the electromagnetic form factor for the $\gamma\Lambda^*\Lambda^*$
vertex.  By varying this cutoff they hoped to be sensitive to a size
effect related to the spatial structure of the \LambdaOne. However, this
$u$-channel effect turned out to be too small in relation to other
theoretical ambiguities to give useful sensitivity.  They used chiral
unitary model results to estimate $g_{K N \Lambda^\ast}$, but left
$g_{K^\ast N\Lambda^\ast}$ as a free parameter.  By comparing to the
very limited experimental data~\cite{Niiyama}, they concluded that
the \LambdaOne{} is produced dominantly by the $s$-channel Born
contribution, without resonant intermediate states, and not by meson
exchange in the $t$-channel.  In an older model, by Williams, Ji and
Cotanch~\cite{Williams:1991tw}, emphasis was placed on constraints from
crossing symmetry and duality, but gauge invariance was not enforced.
The $K^*$ exchange diagram was omitted, and the \LambdaOne{} appeared
via resonance in the $u$-channel.  After satisfactory fits were made to
photoproduction data for the ground state $\Lambda$ and $\Sigma^0$, a
prediction for the threshold cross section of the $\Lambda(1405)$ was
made. These two model cross sections will be compared with our results
later in Sec.~\ref{subsection:results:results1405}.

Photoproduction of the \LambdaOne{} has also been studied by Nacher
\etal~\cite{Nacher:1998mi} in the context of examining the $\Sigma
\pi$ line shapes. The study used an energy- and angle-independent
Weinberg-Tomozawa contact interaction, implying a featureless
differential cross section. So while that study was crucial to our
previous work in Ref.~\cite{lineshapepaper}, it is not relevant for
the current work of examining the cross sections.

\section{Experimental Setup and Data Analysis}
\label{section:SetupAndAnalysis}

The data for this experiment were obtained with the CLAS detector,
located in Hall B at the Thomas Jefferson National Accelerator
Facility, during May and June of 2004. The run, known as g11a, used a
40-cm unpolarized liquid hydrogen target and an incoming unpolarized
real photon beam. Bremsstrahlung photons were created via the CEBAF
accelerator electron beam and a thin gold foil radiator, with an
endpoint energy of $4.019$ GeV.  Electrons that radiated a photon were
detected with the CLAS tagger~\cite{Sober} to obtain energy and timing
information over the range from 20\% to 95\% of the endpoint energy.
Discussion of the CLAS apparatus can be found in Ref.~\cite{CLAS-NIM}.
All aspects of the data collection and initial data handling were
discussed in our previous paper~\cite{lineshapepaper} that focused on
the invariant mass distributions or ``line shapes'' of the
$\Lambda(1405)$. The same analysis extracted the cross sections
presented here. Here we review and discuss the important steps related
to obtaining the cross sections.

The exclusive channels were reconstruction from the detected $K^+$ and
all but one of the hyperon decay products. A one-constraint kinematic
fit to the missing particle was used to select each of the channels;
for the decays $\Lambda(1405)/\Lambda(1520) \to \Sigma^0\pi^0$ both a
photon and a $\pi^0$ were missing, so a simpler missing-mass selection
was used instead.  The hyperon yields for a given decay channel were
obtained in each photon energy and kaon angle bin using an incoherent
fit of MC simulations of signal and background channels. The energy
bins in center-of-mass energy $W$ were $100$ MeV wide.  The angle bins
in the kaon angle \costhetakp{} were again in the center-of-mass
system. In all cases, the fits to each energy and angle bin were
treated independently. We next give some examples to illustrate this
process.


\subsection{Yields for $\Sigma^{0}(1385)$}
\label{subsection:setup:yields1385}

\begin{figure*}[p!t!b!h!]
  \begin{center}
    \subfloat[$W=2.1$ GeV, $\costhetakp=0.85$]
    {\label{fig:Fit1385:2_18}\includegraphics[width=0.5\textwidth]{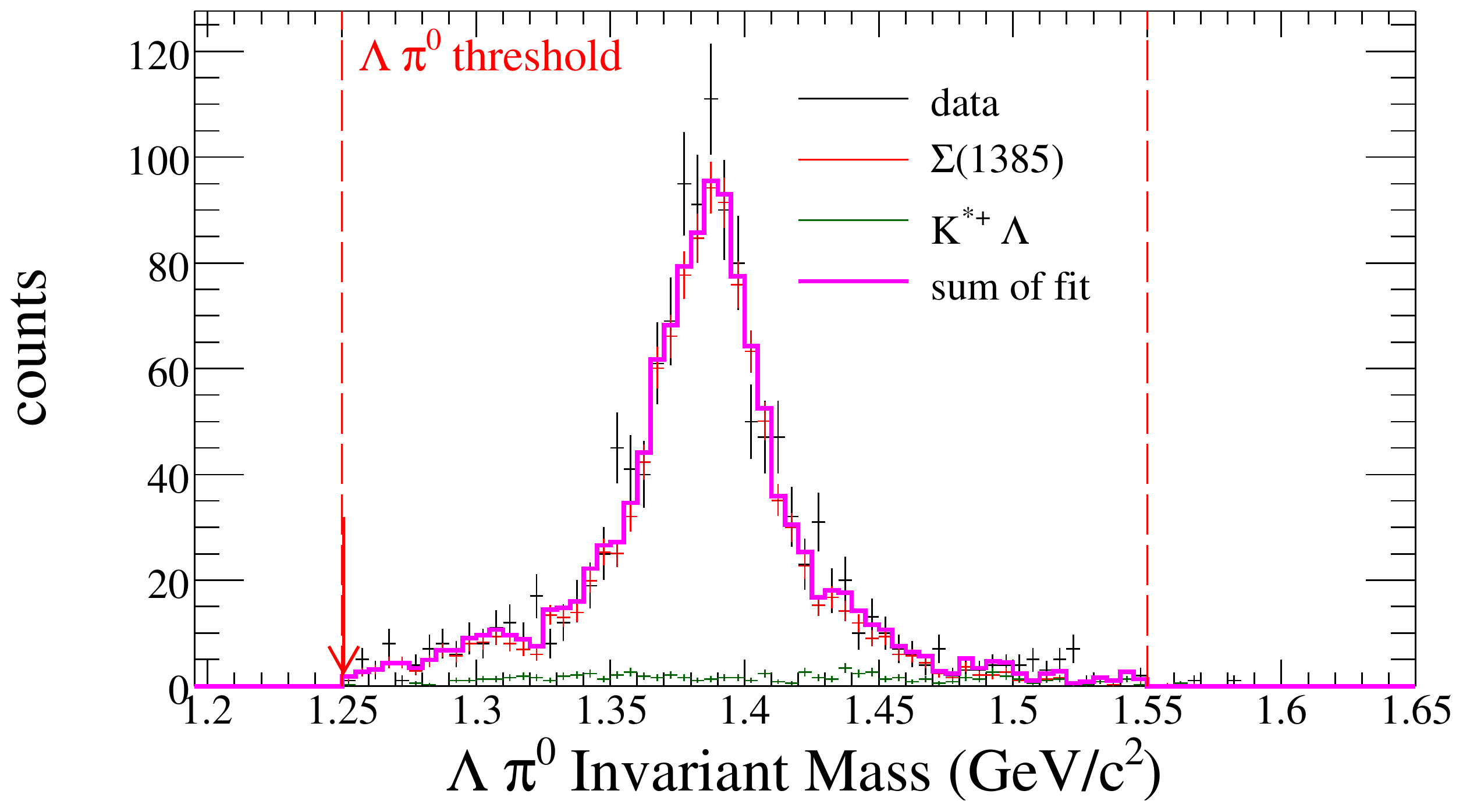}}
    \hfill
    %
    \subfloat[$W=2.6$ GeV, $\costhetakp=0.65$]
    {\label{fig:Fit1385:7_16}\includegraphics[width=0.5\textwidth]{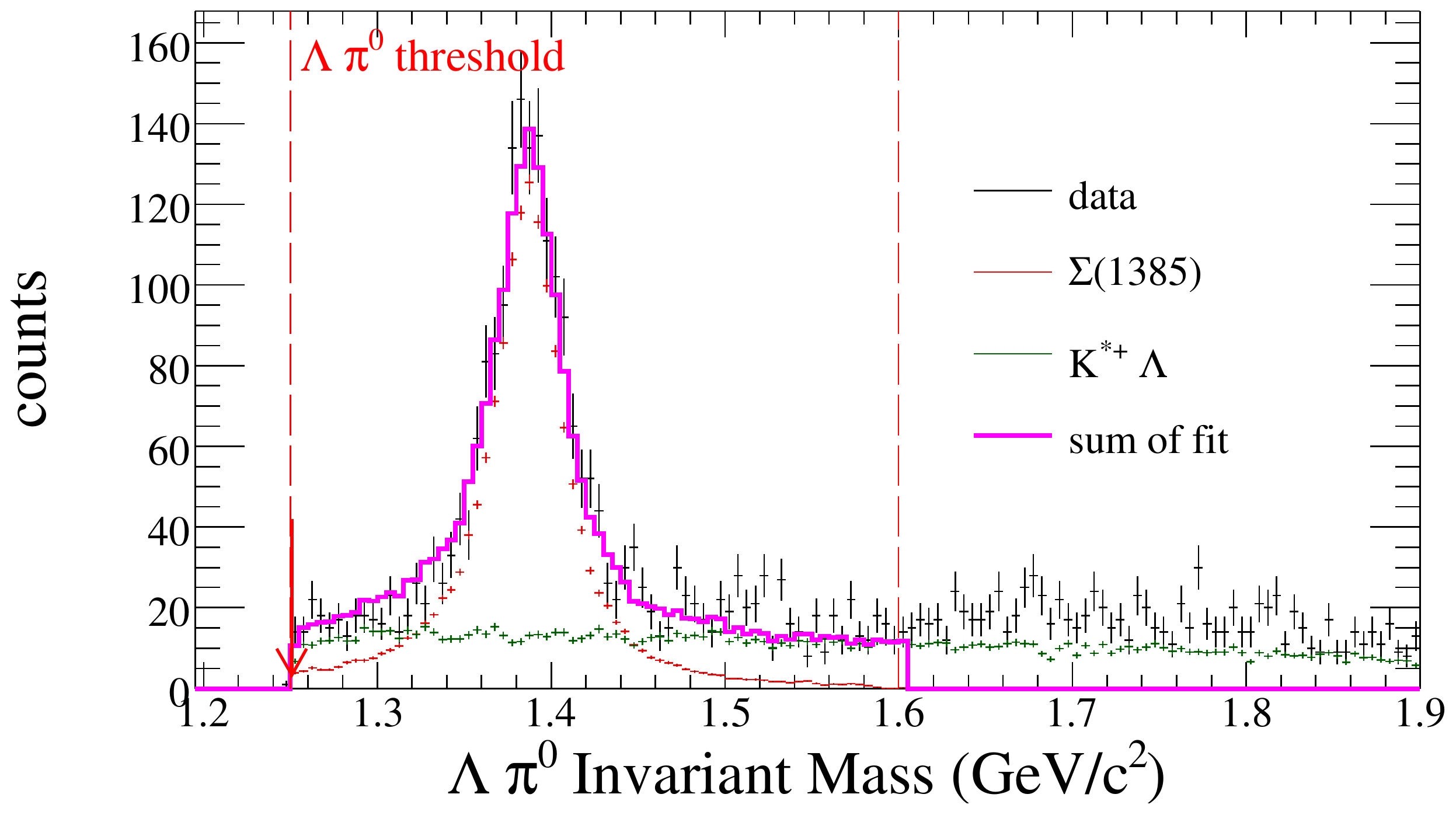}}
    \hfill
    %
    %
    %
  \end{center}
  \caption[Fits of $\Sigma^{0}(1385)$.]{(Color online)
    Sample fit results of the strong
    final state of $\kp \Lambda \pizero$. The events are plotted
    versus the missing mass from the \kp, which is equivalent to the
    invariant mass of the $\Lambda\pizero$ system. The data are shown
    by the black crosses, while the $\Sigma^0(1385)$ signal MC
    and the $K^{\ast +} \Lambda$ background are shown by the red
    crosses and green circles, respectively. The sum of the MC
    templates is shown by the solid magenta line.
    }
  \label{fig:Fit1385}
\end{figure*}

For photoproduction of the $\Sigma^{0}(1385)$, the cross section was
reconstructed from the dominant decay mode to $\Lambda \pi^0$
(B.R. $87.0\%$), where a kinematic fit to the undetected $\pi^{0}$ was
used to optimize the measurements of the energy and momentum of the
detected particles.
Figure~\ref{fig:Fit1385} shows two typical sample bins in
center-of-mass energy bins that are $100$ MeV wide and span a range of
$0.1$ in $\costhetakp$. A full MC simulation of the $K^+
\Sigma^{0}(1385)$ signal reaction was done to create realistic
templates which could be fitted to the data.  Similarly, the dominant
$K^{\ast +} \Lambda$ background channel was simulated.  As discussed
in our previous paper~\cite{lineshapepaper}, the line shape function
for the $\Sigma^{0}(1385)$ was most realistically modeled using a
non-relativistic Breit-Wigner form. This fit the data best in all
energy and angle bins. The $\Sigma^{0}(1385)$ yield was taken to be
the integrated fitted counts in the template line shape.

In some bins in $W$ there was kinematic overlap of the signal hyperon
and the background $K^{\ast +} \Lambda$ events. Tests showed that
there was no discernible coherent interference between the signal and
background channels.  That is, the cross section results did not
change even when a drastic ($\pm 1\Gamma$ or $\sim 100$~MeV) cut was
made to reject events in the kinematic overlap region that contained
most of the $K^{\ast+} \Lambda$ events.

\subsection{Yields for \LambdaOne{} and \LambdaTwo}
\label{subsection:setup:yields1520_1405}

\begin{figure*}[t!b!p!h!]
    \subfloat[]{
      \label{fig:fitcase2SigmaPlus:6_13:nominal}\includegraphics[width=0.48\textwidth]{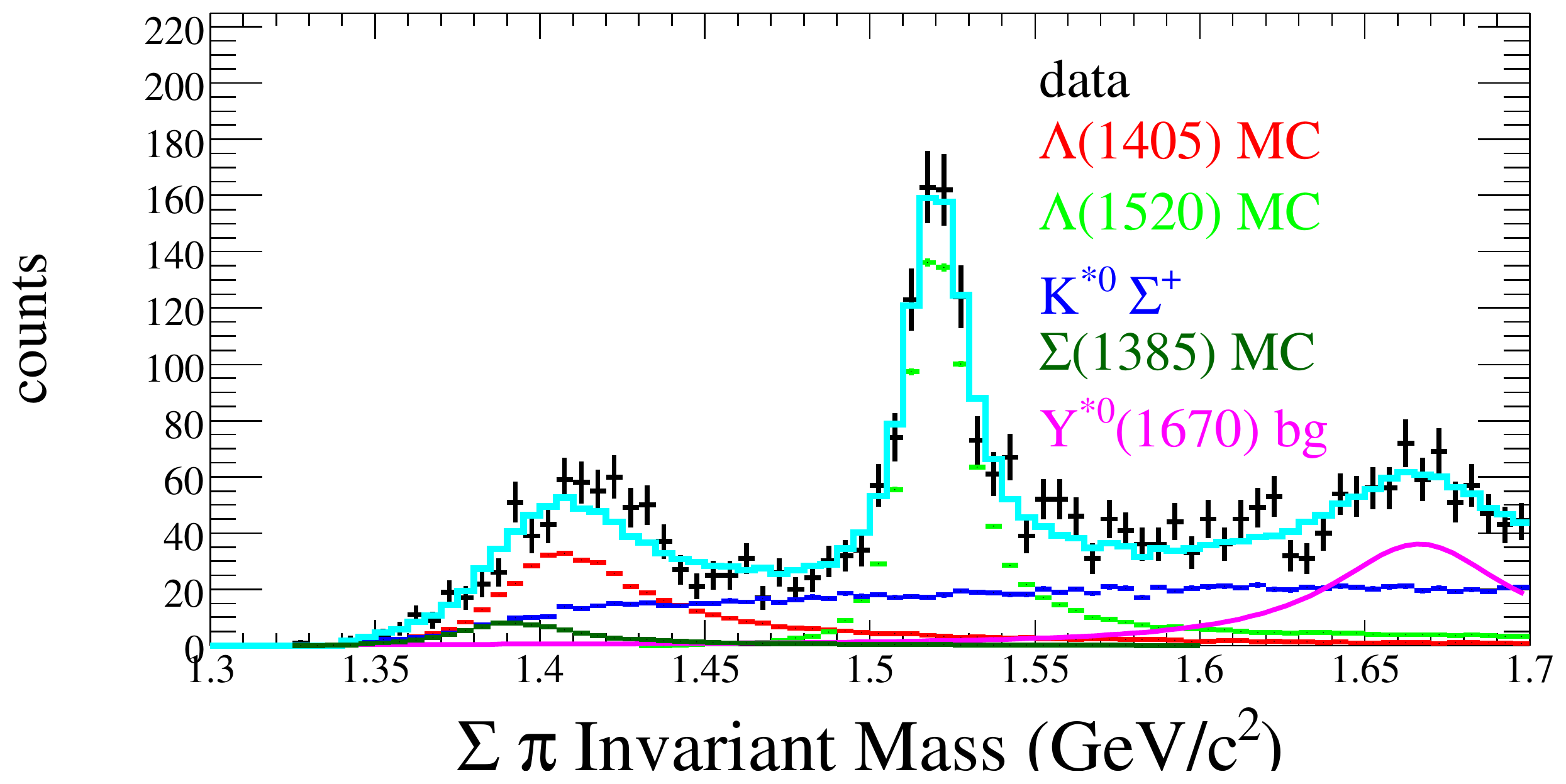}
    }
    \hfill
    \subfloat[]{
      \label{fig:fitcase2SigmaPlus:6_13:dataWithTail}\includegraphics[width=0.48\textwidth]{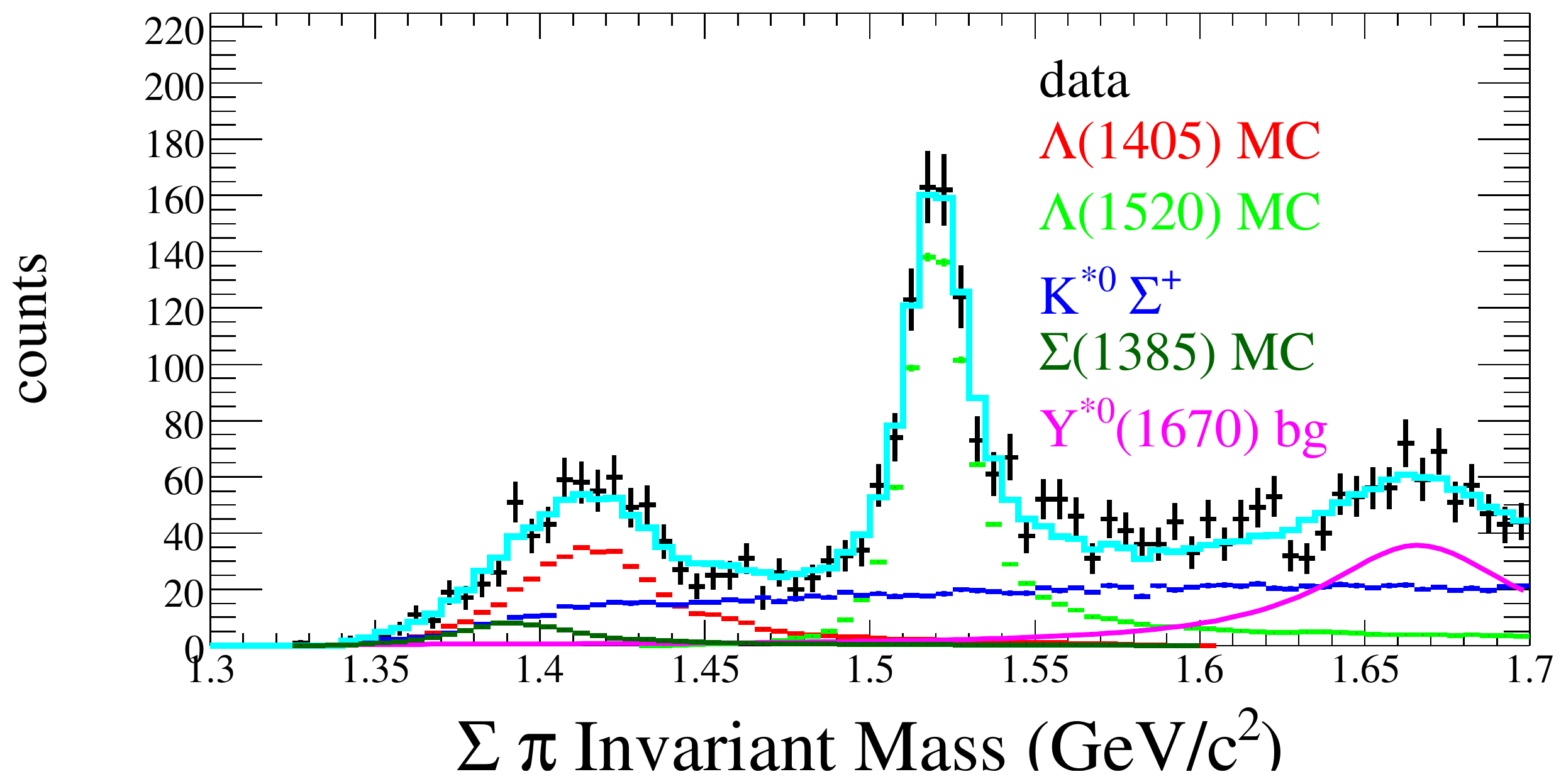}
    }
    \caption{(Color online) Fit result to the strong final state of
    $ \SigmaPlus\pim$ before MC iteration, for
    $W=2.5$~GeV, $\costhetakp=0.35$. The data are
    the black crosses.  
    \protect\subref{fig:fitcase2SigmaPlus:6_13:nominal} 
     The initial fit.  Each
     MC template is shown by a different color, as well as the
     Breit-Wigner function for the $Y^{\ast}(1670)$. The sum of the fit
     is shown in cyan.  
     \protect\subref{fig:fitcase2SigmaPlus:6_13:dataWithTail} 
     Same as in (a), but after two
     iterations of the \LambdaOne{} template to match the data.
    }
    \label{fig:fitcase2SigmaPlus}
\end{figure*}

For photoproduction of the \LambdaOne{} and the \LambdaTwo, the cross
sections were determined using the $\Sigma\pi$ decay mode (B.R 100\%
and 42\%, respectively).  How this was done is illustrated in
Fig.~\ref{fig:fitcase2SigmaPlus}.  Again, MC simulations were used to
realistically model the distribution of events from each of the signal
reactions.  In the case of the \LambdaTwo{} the template shape was a
relativistic Breit-Wigner using PDG parameters and processed through
the CLAS simulation.  The \LambdaOne{} line shape was determined using
an iterated MC method since it did not conform to a simple
Breit-Wigner shape.  The figure shows the initial and the final
iteration of the sample fit. The final iteration matches the data much
better in the \LambdaOne{} region.  The fit also included three
background processes. The first was photoproduction of $K^{+}
\Sigma^{0}(1385)$, which has a known small branching fraction to
$\Sigma\pi$. Its strength therefore was not allowed to float in the
fit, since the dominant $\Lambda\pi^0$ decay mode determines its size
in the $\Sigma \pi$ decay modes, after being corrected by the ratio of
known branching fractions and acceptance in each channel.  The second
backgrounds were the $K^{\ast 0} \Sigma^+$ channels that appeared as a
broad distribution in the example shown.  Finally there was
photoproduction of a higher-mass $Y^\ast$ centered around $1670$
\mevcc, which was modeled with a simple Breit-Wigner line shape and
not with the full MC simulation, as can be discerned from the smooth
curve in Fig.~\ref{fig:fitcase2SigmaPlus}.

The important difference between the results for the \LambdaOne{} and
\LambdaTwo{} is that the particle yields for the former depended
significantly on the final charge state ($\Sigma^+\pi^-$,
$\Sigma^0\pi^0$, $\Sigma^-\pi^+$), while those for the latter did not,
as will be seen later.  When integrating over all production angles
\costhetakp, the mass distribution or ``line shapes'' in the charged
decay modes of the \LambdaOne{} differed, as detailed in
Ref.~\cite{lineshapepaper}.  In contrast, for this paper we take the
summed yields across each $\Sigma \pi$ mass distribution and present
the differential cross sections as a function of production angle.

Tests were carried out to verify that the $K^{\ast 0} \Sigma^+$
background did not cause changes in the cross section or the line
shapes of the hyperons.  Indeed, no interference effects were seen in
the photoproduction line shapes~\cite{lineshapepaper} or in the
extracted differential cross sections, so that the incoherent sum used
here was justified.

\subsection{Acceptance and normalization}
\label{subsection:setup:acceptancenormalization}

A large number of MC events were processed using the GEANT-based
standard CLAS simulation package GSIM.  The event generator used a
bremsstrahlung photon energy distribution.  The kaon production angle
distributions were taken to have an empirical $t$-slope matching the
one for ground state $K^{+} \Lambda$
photoproduction~\cite{Bradford_xsec}.  All (quasi-) two-body decays
were taken to be isotropic in their own rest frames. The generated
events were passed through the detector simulation, and particle
momenta were smeared to match the actual data. In an earlier detailed
analysis of the g11a data set~\cite{MW-thesis}, it was found that the
trigger conditions for this run were not ideally simulated, so a
momentum-dependent \textit{ad hoc} trigger efficiency correction of
$\sim 5\%$ was applied. After all corrections were made, the simulated
events were passed through the same analysis procedures as the actual
data. A further small correction was applied for events with a
$\Lambda$ in the strong final state that compensated for decays that
occurred outside the CLAS time-of-flight Start Counter, as described
in Ref.~\cite{lineshapepaper}.

The photon flux in each energy bin was determined so that the
differential cross sections could be computed. This was done using the
CLAS-standard method based on counting out-of-time electrons in the
photon tagger within well-defined time windows. Also, a correction was
made for the measured $\simeq 70\%$ transmission of photons from the
tagger, through collimators, to the physics target.  Other corrections
were made to handle photon tagger counters not in the primary trigger,
and for the overall measured $\sim$$85\%$ DAQ live-time for this data
set.  Thus, the cross sections reported here are absolutely
normalized.  As mentioned previously, other published CLAS data sets
used the same normalization procedures and have the same level of
normalization uncertainty.

\subsection{Systematic uncertainties}
\label{subsection:setup:systematics}

There were overall or global systematic uncertainties from the yield
extraction and acceptance, flux normalization, and the line shape
fitting procedures. These are discussed in Ref.~\cite{lineshapepaper}
and at more length in Ref.~\cite{Moriya-thesis}.  For the final
systematic uncertainty, the most typical global uncertainties were
added in quadrature to yield a final value of $11.6\%$. A summary of
each uncertainty is shown in Table~\ref{tab:sys_uncertainties}.  The
largest single contribution was from the overall flux normalization.
This was monitored on an hour-by-hour basis by measuring the $\omega$
production yield~\cite{Williams2}, and the uncertainty for the
normalization was determined to be $7.3\%$.

\begin{table}
  \caption{\label{tab:sys_uncertainties} The global systematic
    uncertainties in the experiment. They can be divided into yield
    extraction, acceptance, target characteristics, photon flux
    normalization, and branching ratios~\cite{Beringer:1900zz}. The
    total was calculated by adding the most typical values in quadrature.}
    \begin{tabular}{l r}
      \hline 
      \hline
      Source           & Value ($\%$) \\ \hline
      \deltaTOF{} cuts & $2$--$6$ \\
      Confidence level on kinematic fit & $3$--$12$ \\
      Selection of intermediate hyperons & $2$--$3$ \\
      Target density & $0.11$ \\
      Target length  & $0.125$ \\
      Photon normalization & $7.3$ \\
      Live-time correction & $3$ \\
      Photon transmission efficiency & $0.5$ \\
      $\Sigma^{0}(1385) \to \Sigma \pi, \Lambda \pi$ & $1.5$ \\
      $\Lambda \to \proton \pim$ & $0.5$ \\
      $\SigmaPlus \to \proton \pizero, \neutron \pip$ & $0.30$ \\
      $\SigmaMinus \to \neutron \pim$ & $0.005$ \\ \hline
      Total & 11.6 \\
      \hline 
      \hline
    \end{tabular}
\end{table}


\section{Results}
\label{section:results}

\subsection{Results for $\Sigma^{0}(1385)$}
\label{subsection:results:results1385}

\begin{figure*}[h!t!p!b!]
  \includegraphics[width=0.85\textwidth]{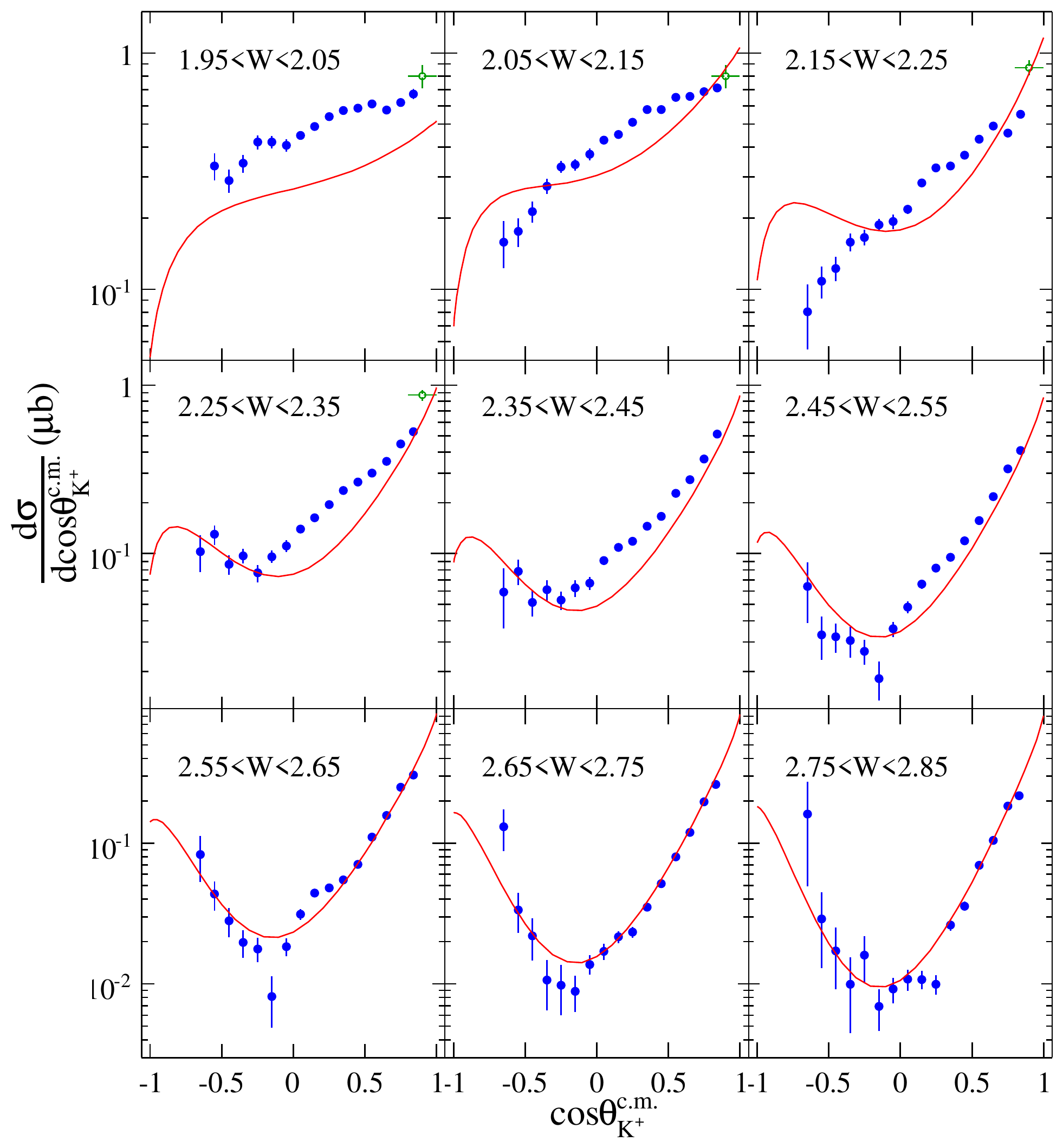}
  \caption{ (Color online) Differential cross section for $\gamma p
    \to K^+ \Sigma^0(1385)$ for the indicated 100 MeV wide bins in $W$
    (GeV).  The CLAS results are the filled blue points, with error
    bars that show the combined statistical and signal channel fit
    (see Fig.~\ref{fig:Fit1385}) uncertainties.  The green hollow
    points are from LEPS~\cite{Niiyama}. The solid red lines are the
    model prediction of Ref.~\cite{Oh:2007jd}.
    }
  \label{fig:xsec1385_all}
\end{figure*}

The $\Sigma^{0}(1385)$ differential cross section was determined
through its dominant decay mode to $\Lambda\pi^0$, as discussed above,
and corrected by the PDG branching fraction of $87\%$ to arrive at the
final results.  The differential cross sections in 100-MeV-wide bins
of $W$ are shown in Fig.~\ref{fig:xsec1385_all} as a function of
center-of-mass kaon angles. They are forward peaked with a fairly
smooth fall-off as a function of angle, but also with a moderate rise
in the backward direction at higher energies. These are the hallmarks
of $t$-channel dominated meson exchange in the production mechanism,
with a hint of $u$-channel baryon exchange to account for the
back-angle rise.  The red curves shown are the calculation of
Oh~\cite{Oh_pc} based on the model of Ref.~\cite{Oh:2007jd}. In
general, the calculation matches the data extremely well at the higher
energies, but undershoots the mid-angles at center-of-mass energies
between $2.25$ and $2.55$ GeV. This may suggest that for
$\Sigma^{0}(1385)$ photoproduction, besides the dominant $t$-channel
exchange mechanism, there may be some resonant contribution.

The model of Ref.~\cite{Oh:2007jd} is essentially a prediction based
only on a preliminary total cross section measurement. In that model,
several $N^{\ast}$ resonances with one- and two-star ratings were
included to produce the ``peak'' in the cross section near $E_{\gamma}
= 2.0$ GeV. Our new differential cross section results should allow
refinement of these estimates.

The only other data available are two data points from the LEPS
Collaboration~\cite{Niiyama} at SPring-8, shown as green hollow
circles at forward angles in the four lowest $W$ bins.  These two
points are plotted twice each since they were obtained in wide energy
bins of $1.5 n< E_\gamma < 2.0$~GeV and $2.0 < E_\gamma < 2.4$~GeV. The
LEPS data were taken at more forward angles than our measurements, and
while the lower energy data is in fair agreement with our
measurements, the higher energy point seems to be somewhat higher than
expected from extrapolation of our results.

\begin{figure}[htpb]
  \includegraphics[width=0.50\textwidth]{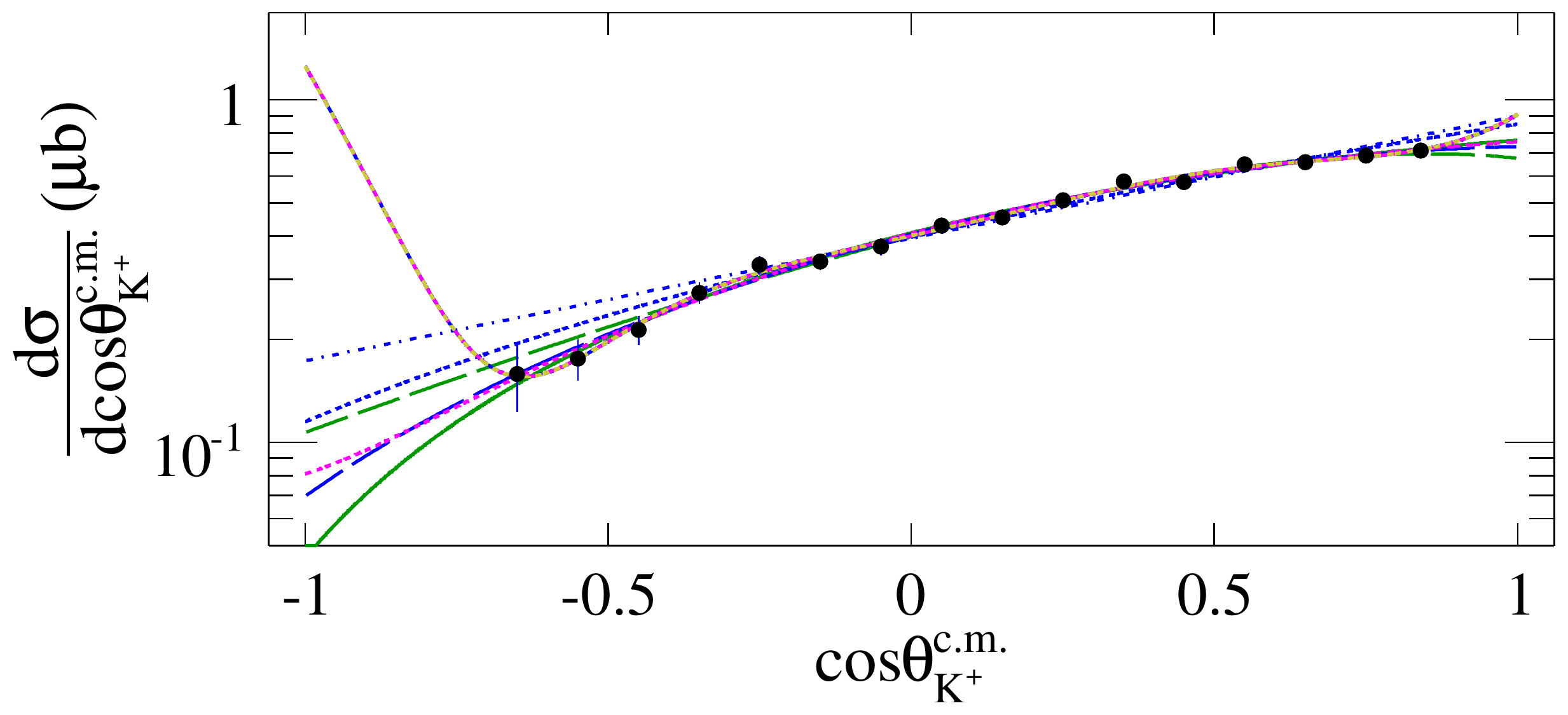}
  \caption{ (Color online) Differential cross section for $\gamma p
    \to K^+ \Sigma^0(1385)$ with possible extrapolations in a single
    $W$ bin for $2.05\leq W\leq 2.15$ GeV.  Some of the nine curves lie on top
    of each other. See text for discussion.}
  \label{fig:xsec1385_fits}
\end{figure}

Figure~\ref{fig:xsec1385_fits} shows the differential cross section in
just one bin of $W$.  The acceptance of CLAS is such that there are
holes in the forward and the backward directions.  In order to
estimate the total cross section an extrapolation into these regions
was necessary.  In the absence of a trusted theoretical model for this
purpose, we averaged an array of plausible functions. The multiple
lines in this figure show the \textit{ad hoc} fits that were performed
for this purpose.  These were made in a purely empirical way in order
to estimate how much variation can reasonably exist, subject to the
constraint that the fits remained positive-definite and that the
extrapolations were not unreasonably wild.

\begin{figure}[htpb]
  \includegraphics[width=0.5\textwidth]{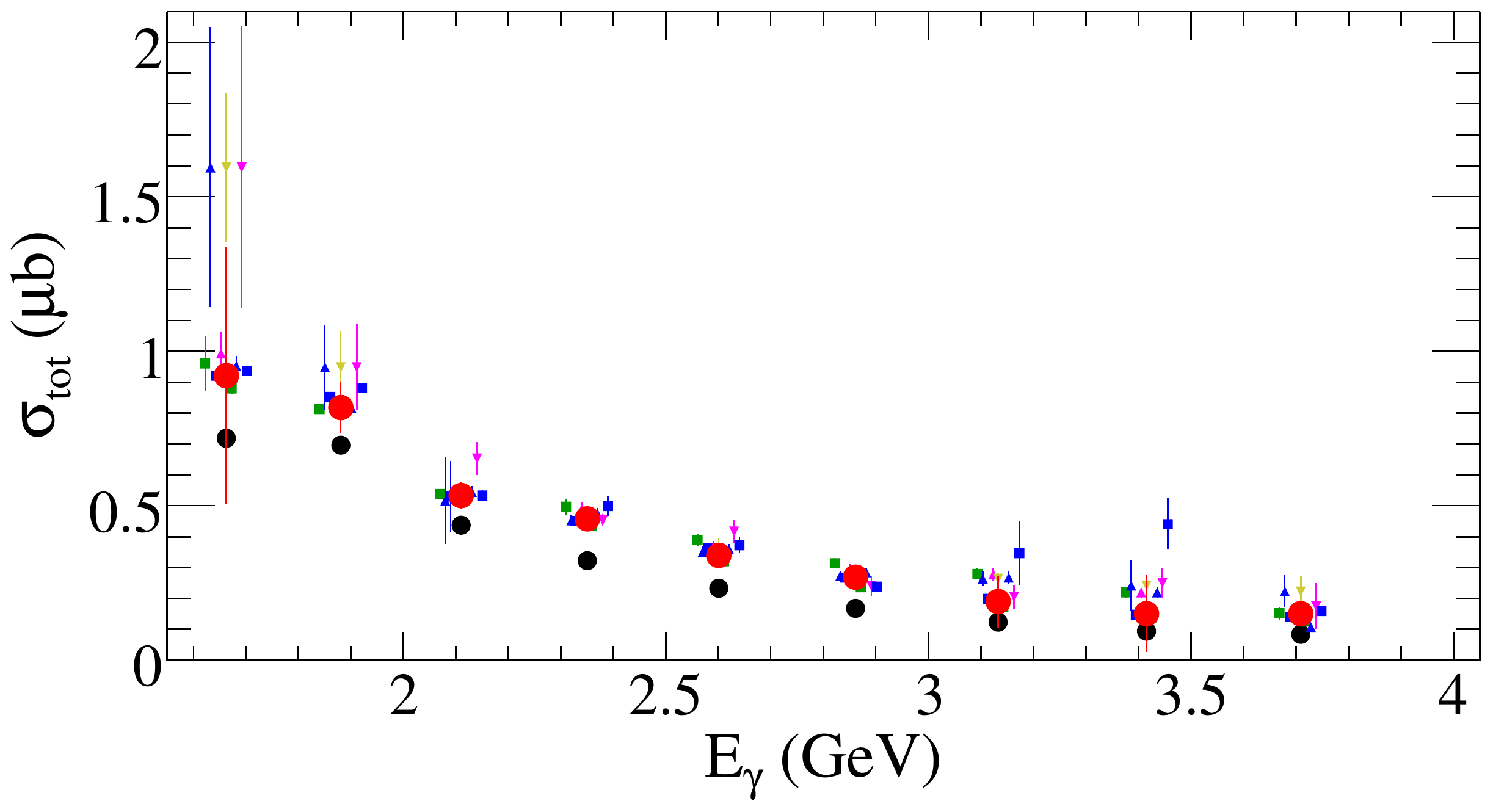}
  \caption{ (Color online) Total cross section of $\Sigma^{0}(1385)$
  as a function of $E_{\gamma}$, showing at each $E_{\gamma}$ the
  collection of alternative extrapolations of the differential cross
  sections tested. The lower black points show the summed
  unextrapolated data.  The large red points show the averaged
  extrapolated values.}
  \label{fig:sigmatot1385_full}
\end{figure}

The functions, using $z = \costhetakp$ were:
\begin{enumerate}
\item \label{itm:sumsquares} absolute square of Legendre polynomials
  $P_{l}(z)$ with complex coefficients $c_{l}$ given in the form of
\begin{align}
  f(z) &= \left|
  \sum_{l=0}^{L} c_{l} P_{l}(z)
  \right|^{2}, \label{eq:fit_base}
\end{align}
with the maximum order $L$ being either $2$ or $3$.
\item the $f(z)$ from Eq.~\eqref{eq:fit_base} with an additive
  exponential of the form $C e^{- D z}$ to account for the forward
  rise. In this case the maximum order used for the Legendre
  polynomials was $L = 1,2,3$.
\item the form of Eq.~\eqref{eq:fit_base}, with a multiplicative 
  overall exponential function $C e^{-D z}$, with $L = 1,2,3$.
\item two exponentials $C_{1} e^{-D_{1} z}$ and $C_{2} e^{+D_{2} z}$
  added to the form of Eq.~\eqref{eq:fit_base}, to take into account
  both the forward and the backward rises. The order used was $L=2$.
\end{enumerate}
Each of these fits was integrated to determine the span of variation
that the total cross section could have.
Figure~\ref{fig:sigmatot1385_full} illustrates how this method of
estimating the total cross section varied depending on the choice of
fit function.  We ascribe no physical interpretation to any of the fit
functions; they merely allowed us to estimate the total cross
sections.  We computed the weighted average of all the individual
points at a given $W$ and assigned that value as the total cross
section, and used the standard deviation of those points as the
estimate of the uncertainty on the total cross sections.

\begin{figure}[htpb]
  \includegraphics[width=0.5\textwidth]{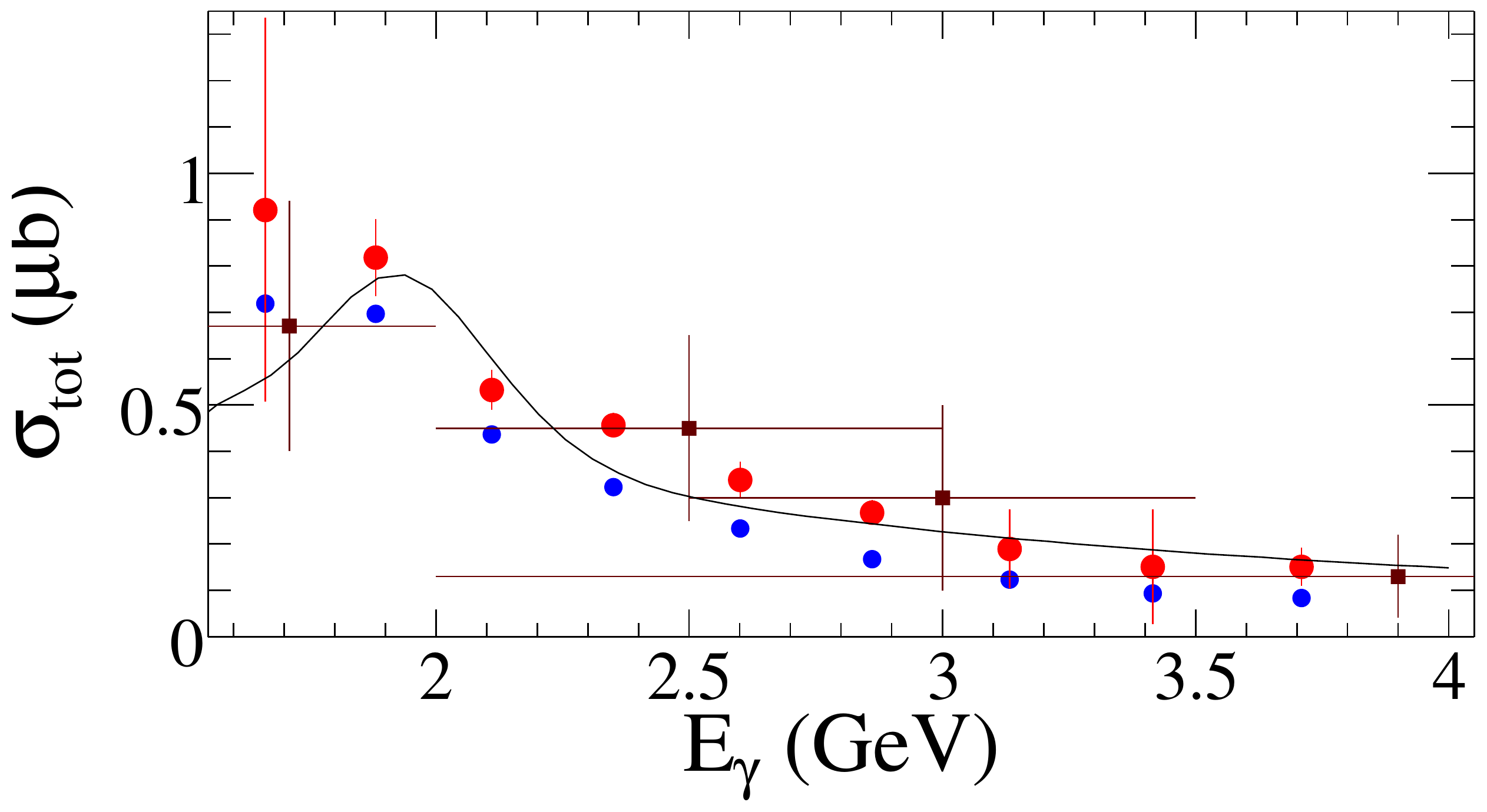}
  \caption{ (Color online) Total cross section for $\gamma p \to K^+
  \Sigma^0(1385)$ as a function of photon energy $E_\gamma$.  The
  small blue points are the summed contribution from the measured
  range of angles in CLAS. The large red points are the extrapolated
  total cross sections using the method discussed in the text.  The
  four dark brown square data points with cross-like error bars are
  bubble chamber data from
  Refs.~\cite{Erbe:1970cq,Erbe:1967,Crouch:1967zz}. The solid black
  curve shows the model of Ref.~\cite{Oh:2007jd}.  }
  \label{fig:sigmatot1385_final}
\end{figure}

The final total cross section result for $\gamma p \to K^+
\Sigma^0(1385)$ is shown in Fig.~\ref{fig:sigmatot1385_final}.  The
small blue points are the summed data across the production angles
accessible in CLAS, while the large red points are the result of the
extrapolations to all angles.  For comparison, we see good agreement
between these results and the much coarser, early bubble chamber data
of Refs.~\cite{Erbe:1970cq,Erbe:1967,Crouch:1967zz}. The solid black
curve is due to the calculation of Oh~\cite{Oh_pc}, and again there is
very good agreement at the higher energies, but there is some
disagreement in the region of $2.0 \leq  E_{\gamma} \leq  2.5$ GeV.

\subsection{Results for \LambdaTwo}
\label{subsection:results:results1520}

\begin{figure}[t!b!p!h!]
  \includegraphics[width=0.50\textwidth]{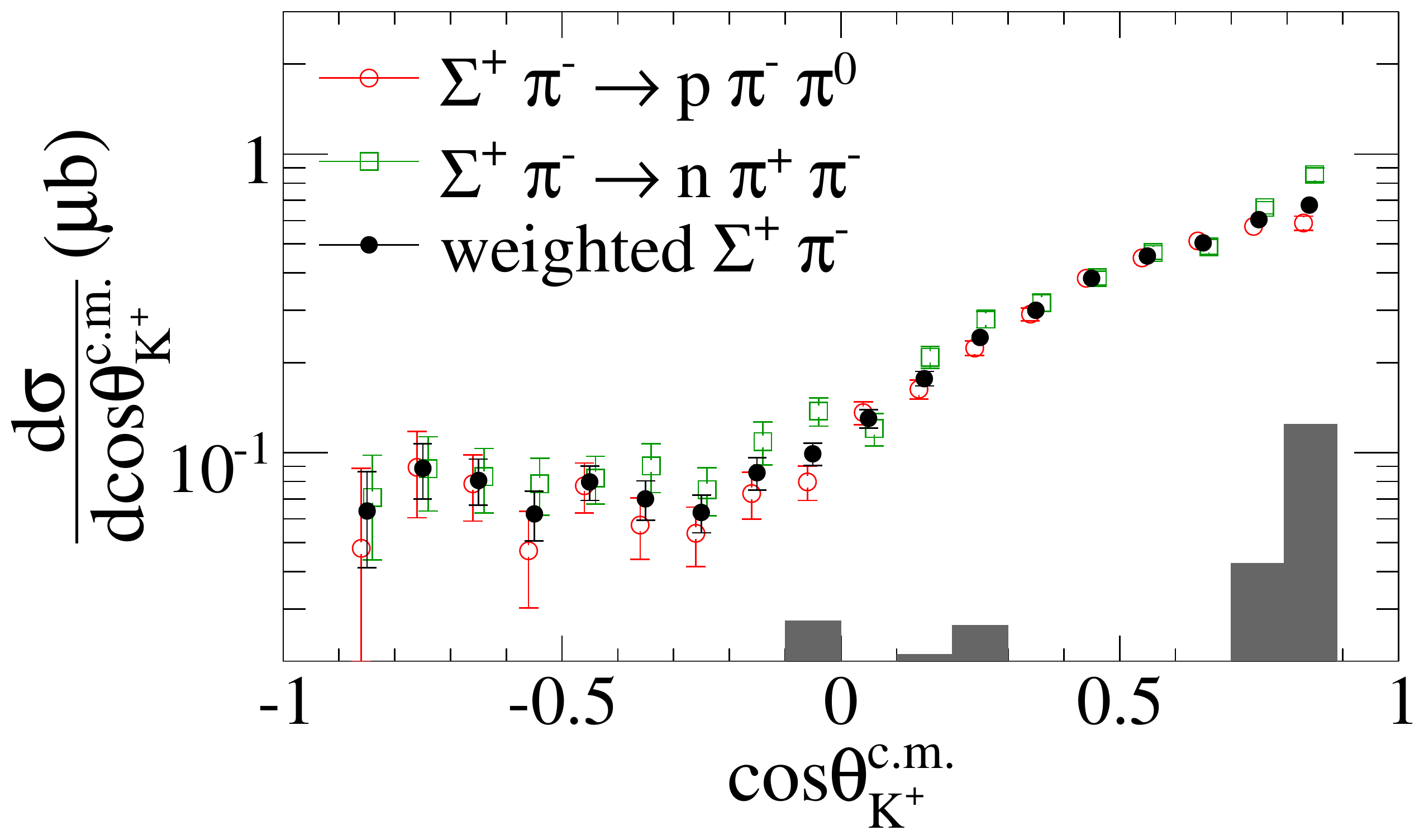}
  \caption[]{(Color online) Comparison of the two $\SigmaPlus \pim$
    channels for the \LambdaTwo{} with $2.45\leq W\leq 2.55$ GeV. The gray
    band is the estimated systematic discrepancy between measurements
    using the two reconstructed final states.  Note the logarithmic
    vertical scale.  }
  \label{fig:xsec1520}
\end{figure}

As discussed in Sec.~\ref{subsection:setup:yields1520_1405}, the
\LambdaTwo{} yields were obtained using
template fits in each energy and angle bin. All charge combinations,
$\SigmaPlus \pim, \SigmaMinus \pip$, and $\SigmaZero \pizero$ were examined.
For the $\Sigma^+$ decays, the two modes $p \pi^0$ and $n \pi^+$ could
be compared for consistency.  Figure~\ref{fig:xsec1520} shows this
comparison in a typical $W$ bin.  The two channels show good
consistency.  The lower gray band shows the angle-by-angle systematic
discrepancy, computed as the difference between the measured final
states with the summed statistical uncertainties subtracted in
quadrature. This quantity is plotted when the difference between the
data points is larger than the sum of the two error bars.  Note that
the vertical scale is logarithmic.  The following results use the
weighted average of the two $\Sigma^+$ final state measurements.

\begin{figure}[t!b!p!h!]
     \subfloat[]
     {\label{fig:xsec1520:W2}\includegraphics[width=0.48\textwidth]{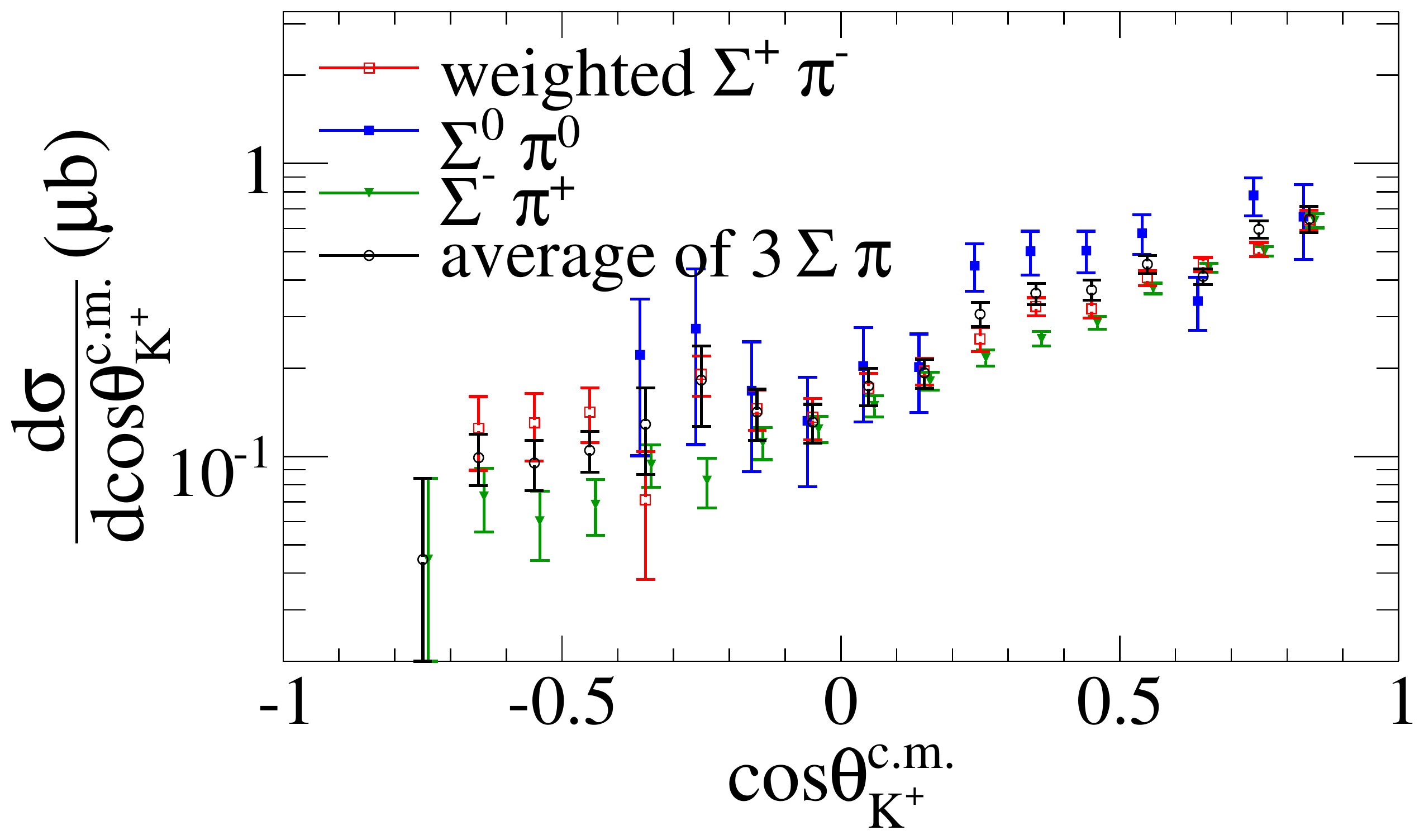}}
     \hfill
     \subfloat[]
     {\label{fig:xsec1520:W6}\includegraphics[width=0.48\textwidth]{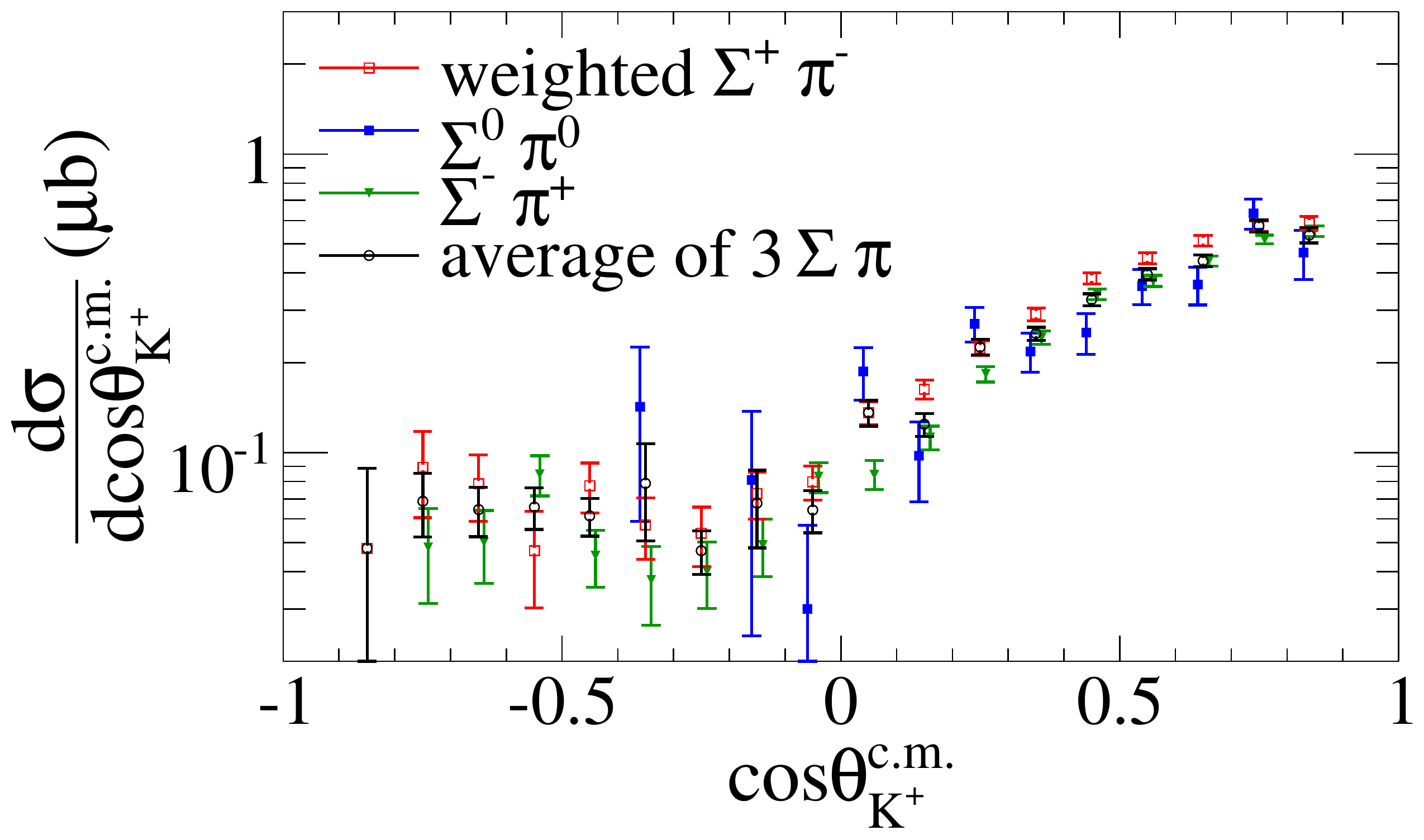}}
     \hfill
  \caption[]{(Color online) Comparison of the three $\Sigma \pi$ decay
    modes for the \LambdaTwo{} in two energy bins at
    \subref{fig:xsec1520:W2} $2.05\leq W\leq 2.15$~GeV and
    \subref{fig:xsec1520:W6} $2.45\leq W\leq 2.55$~GeV as a function of
    center-of-mass kaon angle.  The data points have been shifted
    slightly to prevent overlaps, for clarity.}
  \label{fig:xsec1520_fits}
\end{figure}

Each of the three $\Sigma \pi$ decay modes was reconstructed in the
analysis, so next we compare these three modes for the
\LambdaTwo. This is shown in Fig.~\ref{fig:xsec1520_fits} for two
typical bins in $W$. The agreement among the three decay modes is fair
to good across the range of center-of-mass kaon production angles.  In
principle these should all be identical in the absence of interference
among different isospin contributions to the reaction mechanism. This
was true for the $\Lambda(1520)$, but as shown later, not for the
\LambdaOne, where strong differences are seen depending on the $\Sigma
\pi$ channel.  It was assumed, for the \LambdaTwo{} therefore, that
each decay mode contributes $1/3$ to the total decays to $\Sigma\pi$,
and the measurements scaled accordingly and averaged in each bin over
the measured channels.  The overall branching fraction of $42\%$ for
$\Lambda(1520) \to \Sigma\pi$ was applied to obtain the final values
of the differential cross section for the \LambdaTwo.

\begin{figure}[t!b!p!h!]
%
  \includegraphics[width=0.50\textwidth]{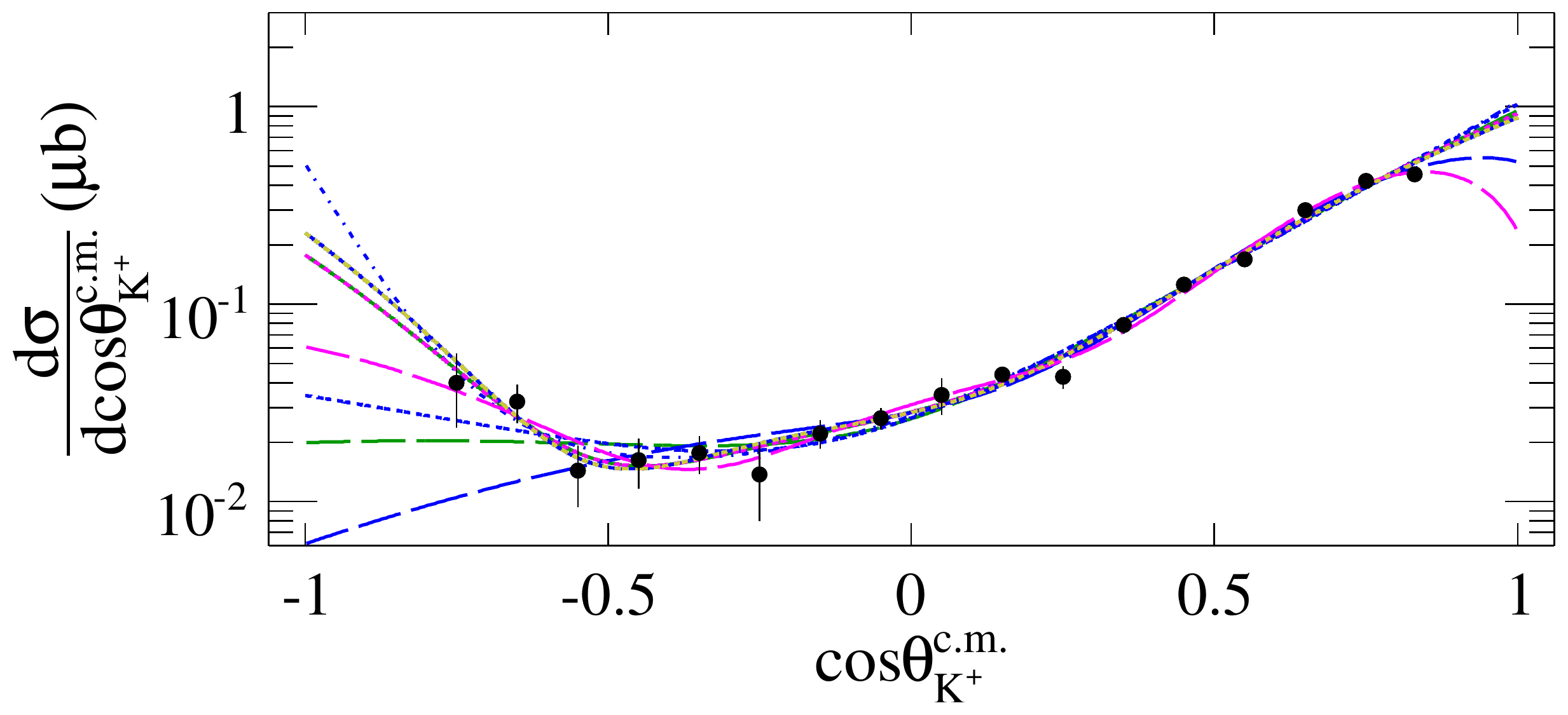}
  \caption[]{(Color online) Differential cross section for $\gamma p
    \to K^+ \Lambda(1520)$ with fit results to a single $W$ bin for
    $2.75\leq W\leq 2.85$ GeV.  The various curves represent possible
    extrapolations to obtain the total cross section.  Some of the
    nine curves lie on top of each other.  }
  \label{fig:xsec1520_cmp}
\end{figure}

The resultant differential cross section for the \LambdaTwo{} in a
single typical $W$ bin is shown in Fig.~\ref{fig:xsec1520_cmp}.  As in
Fig.~\ref{fig:xsec1385_fits}, one sees that the CLAS acceptance did
not extend to the extreme forward and backward kaon production angles,
due to the openings in the spectrometer in those directions.  We
fitted a series of functions in the same manner as done previously for
the $\Sigma^{0}(1385)$ discussed in
Section~\ref{subsection:results:results1385}.  We took the weighted
average of all these variants and their standard deviation as the best
estimated uncertainty for the total cross section.

\begin{figure*}[htpb]
  \includegraphics[width=0.85\textwidth]{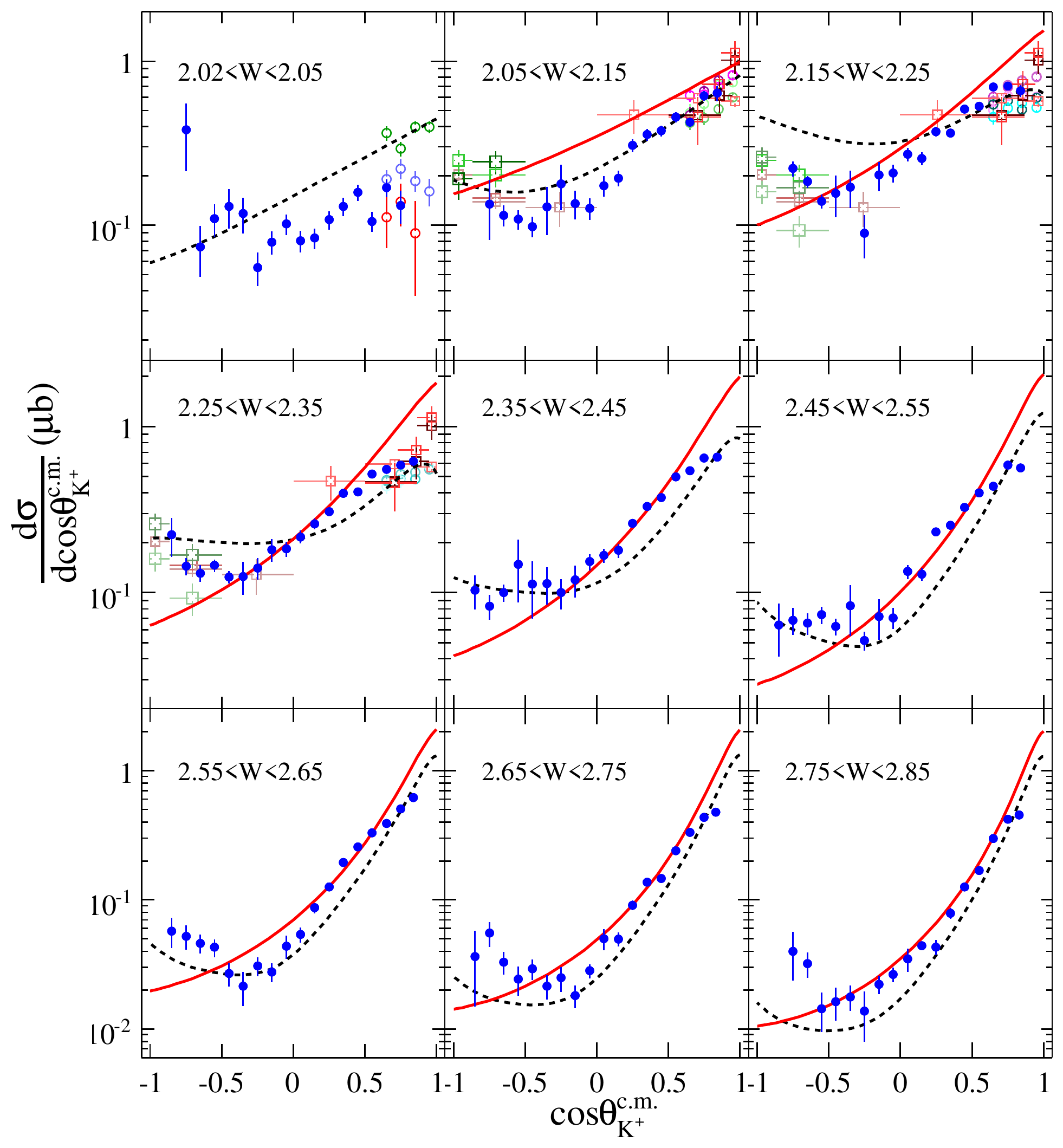}
  \caption{ (Color online) Differential cross section for $\gamma p
    \to K^+ \Lambda(1520)$ for nine bins in center-of-mass energy $W$
    (GeV). The solid blue circular points are the measured CLAS
    data. Measurements from LEPS are shown by the various colored
    hollow square points~\cite{Muramatsu:2009zp}.  The hollow circular
    points show separate LEPS results~\cite{Kohri:2009xe}; in the lowest
    $W$ bin they are for $W=2.01$~GeV (red), $2.03$~GeV (blue), and
    $2.05$~GeV (green).  The lowest $W$ bin is 30 MeV wide, while the
    others are all 100 MeV wide.  The black dashed curve is the model
    calculation by He \etal~\cite{JunHe}, while the red solid curve is
    the model calculation by Nam \etal~\cite{Nam:2010au}. Note that
    the first bin in $W$ is 30 MeV wide while the others are 100 MeV wide.}
  \label{fig:xsec1520_all}
\end{figure*}

The final differential cross sections for the \LambdaTwo{} in all bins
of $W$ are shown in Fig.~\ref{fig:xsec1520_all}.  Near threshold the
CLAS cross section is fairly flat.  Due to the \LambdaTwo{} threshold
at 2.013 GeV, the lowest $W$ bin was normalized using the uniform
integrated photon flux between 2.020~GeV and 2.050~GeV, a 30~MeV
wide region. The other bins are 100~MeV wide.  There is an additional
estimated $\pm36\%$ systematic scale uncertainty in this $W$ bin due
to the finite width of the hyperon, acceptance, and threshold effects.
In the highest energy bin the cross section is quite forward peaked
with a hint of plateauing toward the most forward angles.  Also
evident is that the cross section flattens or even rises slightly
toward large angles.  These are the qualitative hallmarks of
$t$-channel dominance with at least two poles/trajectories at low $-t$
and possible $s-$ or $u$-channel baryon exchange at large angles.

For comparison with the CLAS results, the data from LEPS are
shown. Ref.~\cite{Muramatsu:2009zp} measured the differential cross
section in several final states using various methods that were
presented as equivalently accurate. These points are shown as open
squares in Fig.~\ref{fig:xsec1520_all}.  In the threshold region,
Ref.~\cite{Kohri:2009xe} measured the rapidly-rising differential
cross section with finer energy binning than our present results.
Points that overlap with our bins are shown as the open circles in
Fig.~\ref{fig:xsec1520_all}.  Agreement between CLAS and LEPS is good
or very good across all overlapping bins.

\begin{figure}[htpb]
  \includegraphics[width=0.5\textwidth]{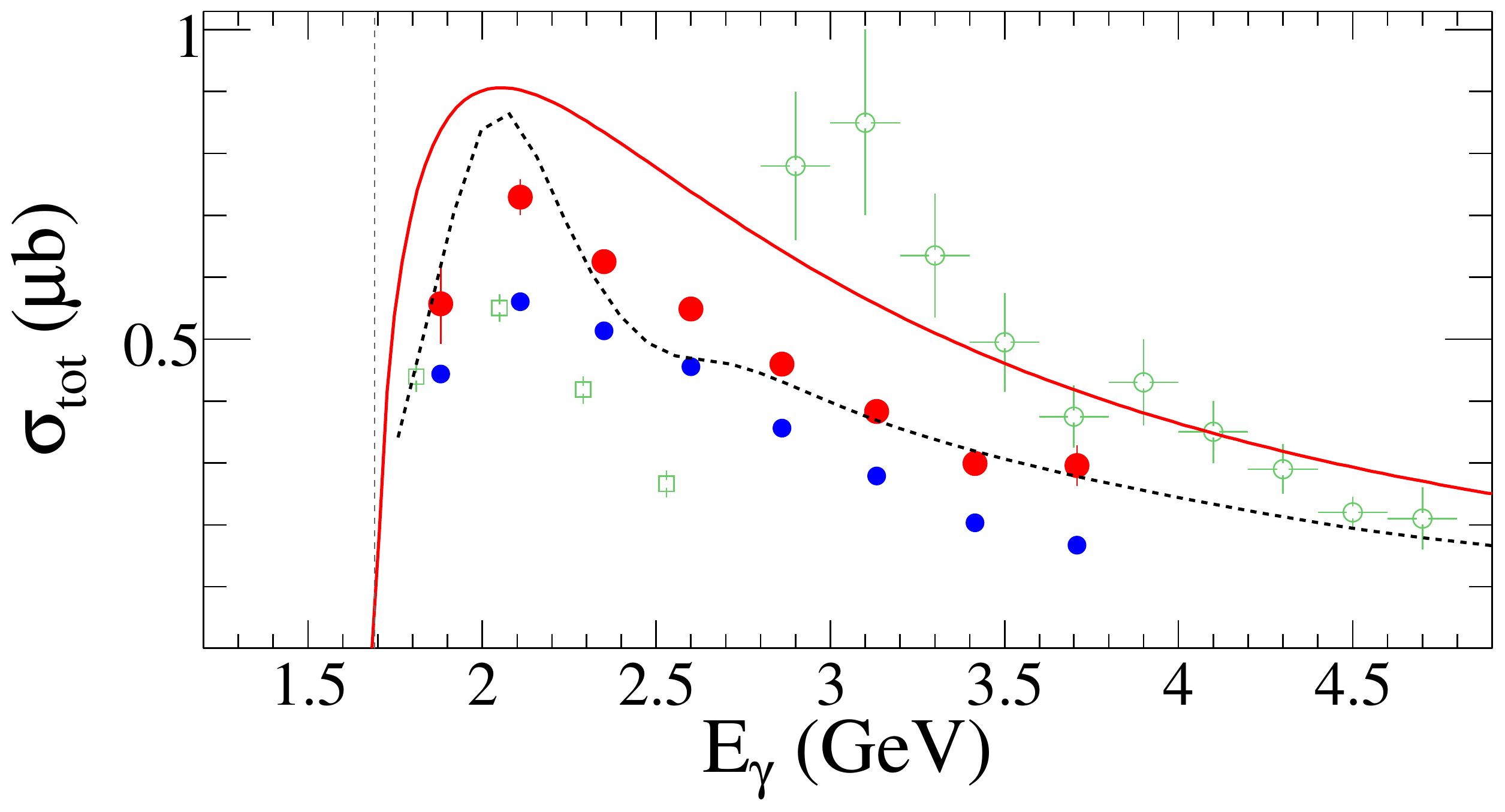}
  \caption{ (Color online) Total cross section of \LambdaTwo{} as a
    function of $E_{\gamma}$. The small blue points are the CLAS data
    summed over the acceptance of the detector, while the larger solid
    red points show the extrapolation over all space.  The error bars
    are the combined statistical and extrapolation uncertainties.  The
    dashed black curve is a calculation by He \etal~\cite{JunHe} and the
    solid red curve is a calculation by Nam \etal~\cite{Nam:2010au}.
    The green hollow squares are measurements by
    SAPHIR~\cite{Wieland:2011zz}, and the green hollow circles are
    measurements from SLAC/LAMP2~\cite{Barber}.}
  \label{fig:sigmatot1520_full}
\end{figure}

The red solid curves are the model of Nam \etal~\cite{Nam:2010au}
computed for the present kinematics~\cite{Nam_pc}.  No Regge
contributions were included, and no $K^*$ exchange in the $t$ channel.
Also, the \LambdaTwo{} anomalous magnetic moment was set to zero,
implying no $u$-channel exchanges.  This leaves the dominant contact
term interaction as well as small contributions from $t$ and $s$
channel Born terms.  Note that this model was developed to match the
scant higher energy data from Ref.~\cite{Barber}.  This is seen in
Fig.~\ref{fig:sigmatot1520_full}, where the red solid curve comes
closest to the Ref.~\cite{Barber} data set but is significantly higher
than the new CLAS results.  This model overestimates the cross section
and lacks sufficient strength at large angles.

The black dashed curves in Fig.~\ref{fig:xsec1520_all} show the model
prediction of He and Chen~\cite{JunHe} computed for the present
kinematics~\cite{He_pc}.  Evidently this model is a closer match to
the new data, but the comparison is not perfect.  It captures the
slight rise at backward angles and is mostly closer in magnitude to
the data.  It also tends to follow the flattening of the total cross
section at forward angles.  This model includes $K^*$ exchange, and
although this is more important at higher energies, it may help
reproduce the forward-angle behavior we see. This model also includes
the $s$-channel $N(2080) D_{13}$ resonance that was quite significant
when matching the data of Kohri~\etal{} from LEPS~\cite{Kohri:2009xe}.
(See also the earlier work in Ref.~\cite{Xie:2010yk}.)  This resonance
is also responsible for the fairly narrow peak in the total cross
section near $2$ GeV shown in Fig.~\ref{fig:sigmatot1520_full}. We can
conclude that this model, using several additional interaction
elements compared to Nam~\etal~\cite{Nam:2010au}, is able to come
closer to reproducing the present differential and total cross section
data for the \LambdaTwo.

Figure~\ref{fig:sigmatot1520_full} shows the total photoproduction
cross section for the \LambdaTwo. The recent SAPHIR
data~\cite{Wieland:2011zz} is in rather strong disagreement with the
data from SLAC/LAMP2~\cite{Barber}.  The new CLAS results lie almost
exactly between these two measurements.  The small blue data points
show the CLAS data summed over the useful acceptance of the detector,
and the larger solid red points show the extrapolated total cross
section.  The model curves correspond to those of
Fig.~\ref{fig:xsec1520_all}.

\subsection{Results for \LambdaOne}
\label{subsection:results:results1405}

The case of the \LambdaOne{} is somewhat different from the others
because each of the $\Sigma\pi$ charged final states yields a
different cross section.  In our previous paper~\cite{lineshapepaper}
this was traced to the fact that there is considerable interference of
an $I=1$ $J^P= 1/2^-$ amplitude with the $I=0$ $J^P= 1/2^-$ amplitude
that, by definition, represents the \LambdaOne.  In
Refs.~\cite{lineshapepaper} and \cite{hyp-proceedings}, isospin
decompositions were done in an attempt to separate those amplitudes as
a function of the $\Sigma\pi$ mass and of $W$. In those articles the
data were integrated over all kaon production angles, but in the
present situation we wish to extract the cross section as a function
of kaon production angle while integrating over the $\Sigma \pi$ mass
distributions.  Unfortunately, with the statistics available in this
measurement it was not possible to do the isospin decomposition in
each energy and angle bin separately.  Instead, we present the results
when integrating the $\Sigma \pi$ mass range from threshold up to
$1.5$ GeV.  Thus, we say that this experiment has measured the cross
section in the ``region'' of the \LambdaOne, without explicit
separation of the isospin amplitudes.

\begin{figure}[t!b!p!h!]
  \includegraphics[width=0.50\textwidth]{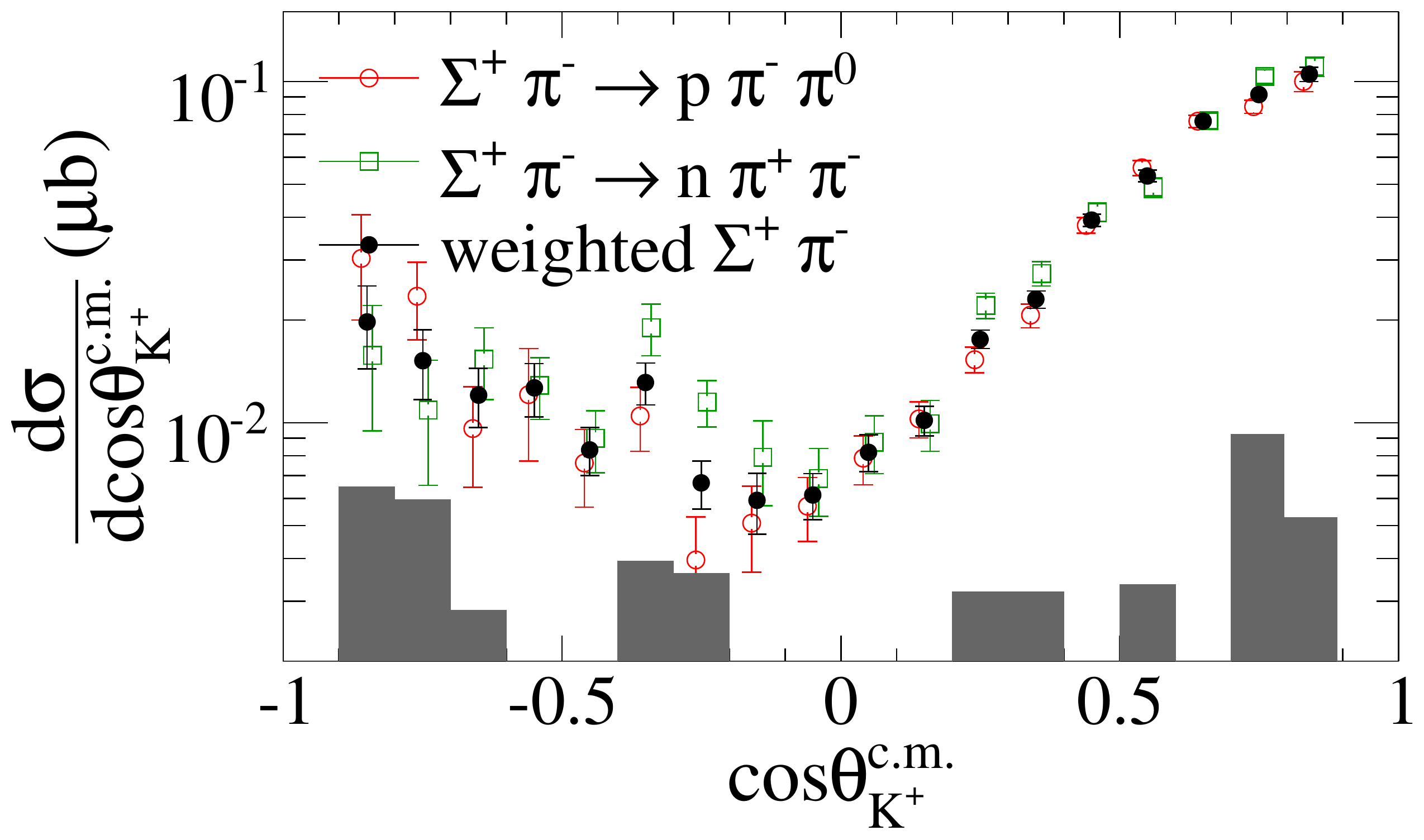}
  \caption[]{(Color online) Comparison of the two $\SigmaPlus$ decay
    modes for the \LambdaOne{} with $2.45\leq W\leq 2.55$ GeV. The gray band
    is the estimated systematic discrepancy between measurements using
    the two reconstructed final states. The points have been shifted
    for visual clarity.  }
  \label{fig:xsec1405}
\end{figure}

As in Sec.~\ref{subsection:results:results1520} for the case of the
\LambdaTwo, there were two decay modes for reconstructing the
$\Sigma^+$.  A sample comparison of these two modes is shown in
Fig.~\ref{fig:xsec1405}.  The agreement between the decay modes was
usually good to very good. In each case, the yield of what we
nominally call the \LambdaOne{} was taken from the line shape
templates in the relevant mass range discussed in
Sec.~\ref{subsection:setup:yields1520_1405}. The gray systematic error
band is the difference between the measured values with the summed
point-to-point uncertainties subtracted off in quadrature. These two
measurements were then averaged together to give the cross section in
the $\Sigma^+\pi^-$ decay mode.

\begin{figure*}[t!b!p!h!]
  \includegraphics[width=0.85\textwidth]{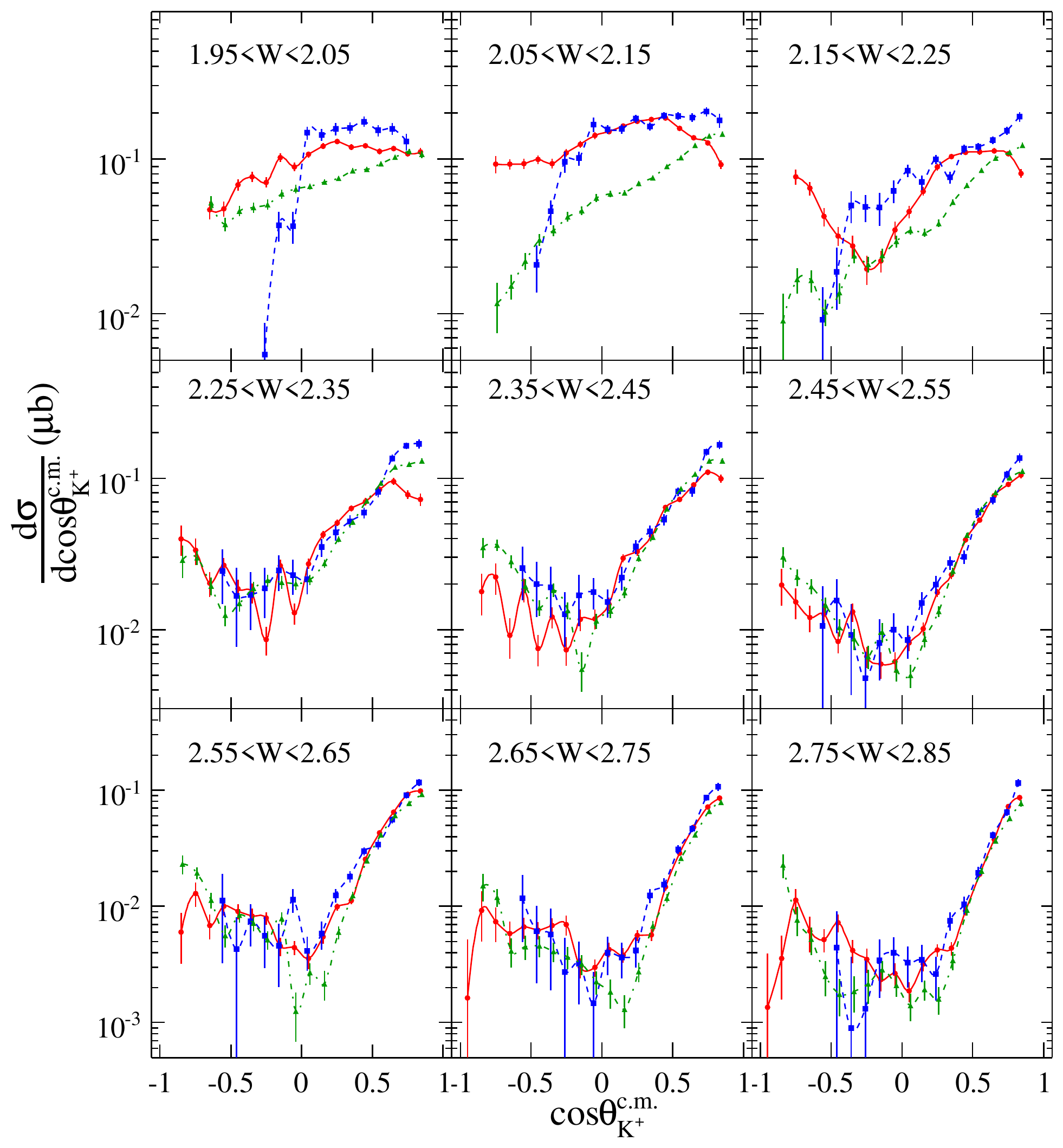}
  \caption[]{(Color online) Differential cross sections for $\gamma p
    \to K^+ Y^*$ in the \LambdaOne{} mass region from threshold to
    $1.5$ \gevcc{} for each of the three $\Sigma \pi$ charge
    states. The ranges of center-of-mass energies $W$ (GeV) are
    indicated. The $\SigmaPlus \pim$ channel is shown as red
    circles/solid line, the $\SigmaZero \pizero$ channel is shown as
    blue squares/dashed line, and the $\SigmaMinus \pip$ channel is
    shown as green triangles/dot-dashed line. The curves are simply
    spline fits to guide the eye.  }
  \label{fig:xsec1405_fits}
\end{figure*}

\begin{figure*}[t!b!p!h!]
  \includegraphics[width=0.85\textwidth]{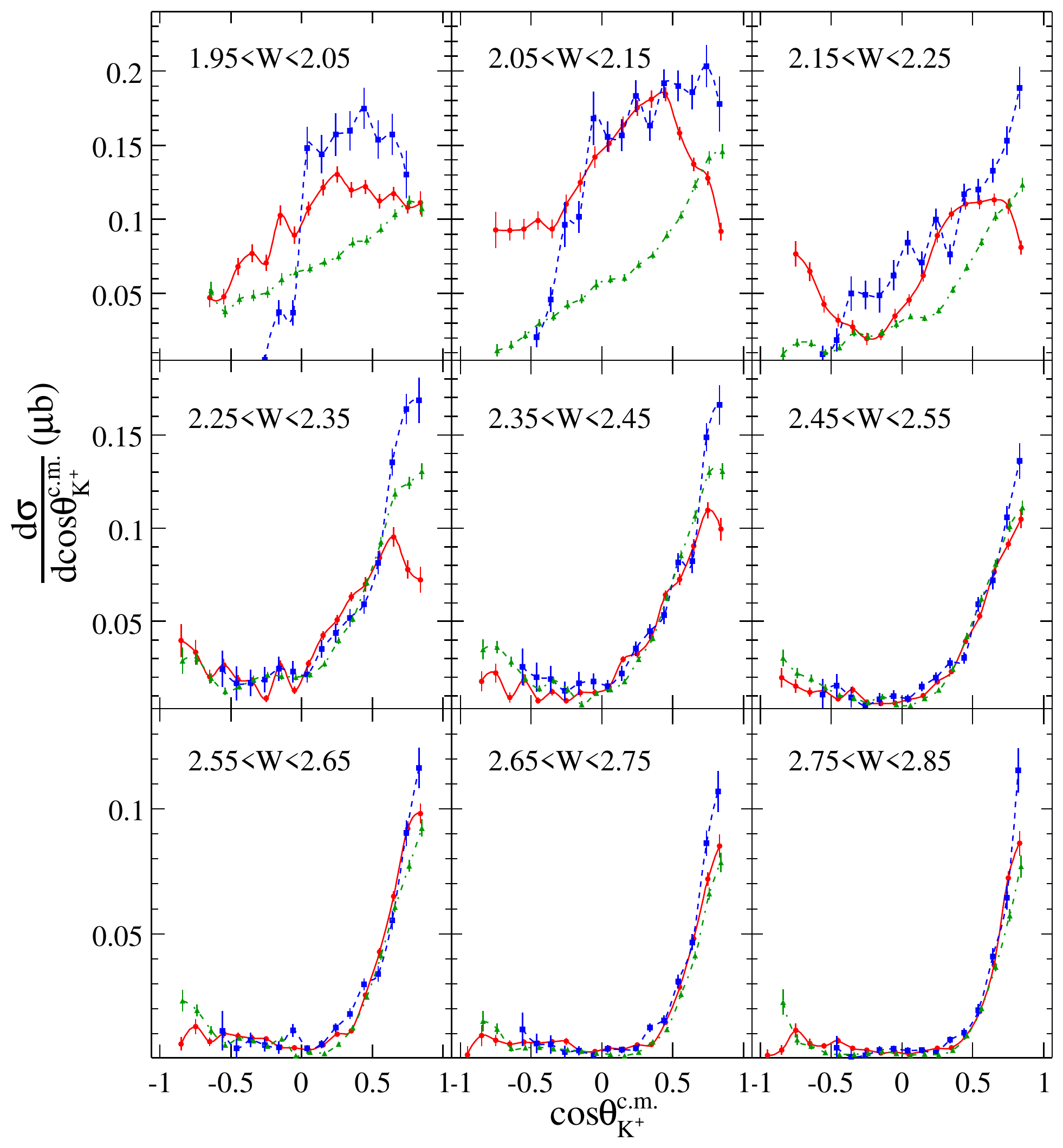}
  \caption[]{(Color online) Differential cross sections for $\gamma p
    \to K^+ Y^*$ in the \LambdaOne{} mass region for each of the three $\Sigma \pi$ charge
    states. The ranges of center-of-mass energies $W$ (GeV) are
    indicated. Same as previous figure, but on a linear vertical scale
    to emphasize the differences in lower $W$.}
  \label{fig:xsec1405_fits_nolog}
\end{figure*}

Comparison of the $\Sigma^\pm\pi^\mp$ and $\Sigma^0\pi^0$ decay modes
of the \LambdaOne{} mass region is shown in
Figs.~\ref{fig:xsec1405_fits} and \ref{fig:xsec1405_fits_nolog} for
all bins in $W$.  In the lower $W$ bins there are very significant
differences between the measured cross sections. Below about $W =
2.3$~GeV there is clearly some additional dynamics present causing the
charge states to have quite different cross sections.  We interpret
this as being due to the interference of $I=0$ and $I=1$ amplitudes in
the production mechanism in the \LambdaOne{} mass range.  The steep
drop in the $\Sigma^0\pi^0$ cross section at large angles for the
lower $W$ bins was checked and found to be correct in this analysis.
Toward the higher end of the $W$ range the three cross sections tend
to merge together into a characteristic forward peak, indicating
$t$-channel dominance in the reaction mechanism.  

At present there is no model calculation available that can explain
this interference effect.  Due to the lower c.m. momenta at lower $W$,
one may speculate that the electromagnetic and hadronic interactions
have more time to develop at lower $W$, hence showing more
interference effects.  Below $W=2.3$~GeV the mean \LambdaOne{} decay
distance is less than the incoming photon wavelength.

To combine these three channels into one effective cross section for
the \LambdaOne{} mass region, we simply added the three components
together.  In Ref.~\cite{lineshapepaper} it was shown that if $t_0$ is
the amplitude for the $I=0$ production and if $t_1$ is the amplitude
for the $I=1$ mechanism that interferes with it, then the sum of the
three decay modes is proportional to $|t_0|^2 + |t_1|^2$, with
cancellation of the interference cross terms.

\begin{figure*}[h!t!p!b!]
  \includegraphics[width=0.85\textwidth]{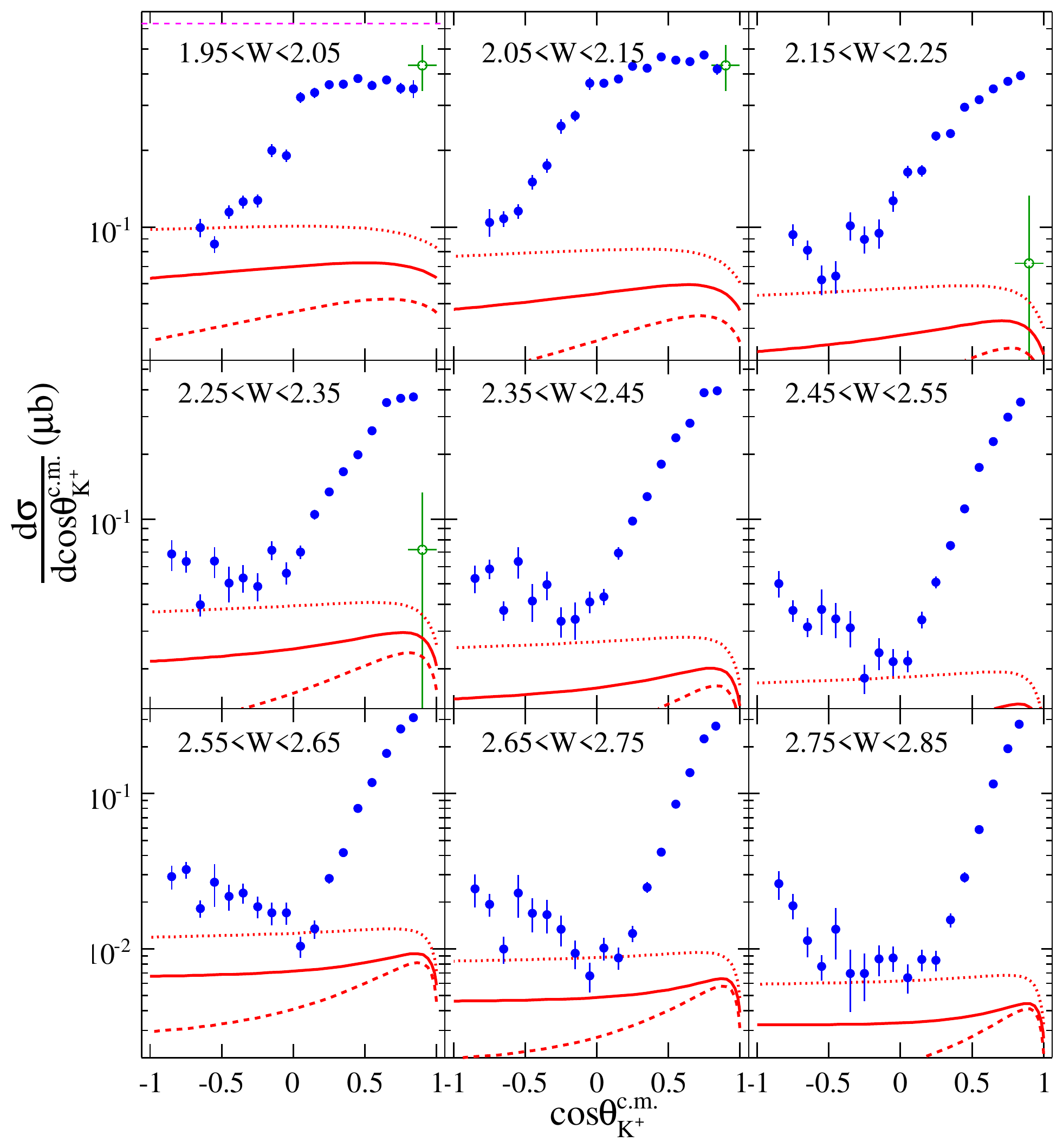}
  \caption{ (Color online) Differential cross sections for $\gamma p
    \to K^+ Y^*$ for the \LambdaOne{} mass region in the indicated 100
    MeV wide bins in $W$ (GeV).  The CLAS data (solid blue points)
    have been summed over the three $\Sigma \pi$ decay modes.  The
    error bars combine statistical and systematic fitting
    uncertainties.  The forward-most datum in the lowest $W$ bin used
    an estimated value for the $\Sigma^0\pi^0$ cross section.  The
    green hollow circle points are from Ref.~\cite{Niiyama}.  The red
    curves are from Ref.~\cite{Nam_pc}, based on the model of
    Ref.~\cite{Nam:2008jy} discussed in the text, while the flat
    dashed magenta line in the $W=2.0$~GeV panel is the prediction
    from Ref.~\cite{Williams:1991tw}.
   }
  \label{fig:xsec_1405}
\end{figure*}

The resulting summed differential cross sections for the \LambdaOne{}
mass range are shown in Fig.~\ref{fig:xsec_1405}. The overall trends
are quite similar to the other two hyperons. We will make a direct
comparison later. The $\Sigma^0\pi^0$ cross section was not measured
in the forward-most angle bin for $W=2.00$~GeV, as seen in
Fig.~\ref{fig:xsec1405_fits}.  The summation used the neighboring value
for this datum, with an error bar enlarged to the size of the
difference between the two neigboring points.

The only data comparison available is the LEPS
Collaboration result by Niiyama~\etal~\cite{Niiyama} in the lowest $W$
bins.  Their two data points are plotted twice each since their energy
bins were quite wide compared to ours. The claim made in that paper
was that the ratio of \LambdaOne{} to $\Sigma^{0}(1385)$ drops from
$0.54\pm0.17$ to $0.084\pm0.076$ between their two energy bins.  The
CLAS results do not support the LEPS observation.  For $W$ between
2.15 and 2.35 GeV the CLAS cross sections are about a factor of six
larger than those of the previous experiment.

A recent prediction for the \LambdaOne{} photoproduction cross section
due to Nam \etal~\cite{Nam:2008jy} is plotted in
Fig.~\ref{fig:xsec_1405}. In their effective Lagrangian model the
$s$-channel Born term is dominant. There is interference with
$K^{\ast}$ exchange, and the three curves stem from letting
$g_{K^{\ast} N \Lambda^{\ast}}$ take the values $0$ (solid red),
$+3.18$ (dotted red), or $-3.18$ (dashed red). Evidently, this model
omits a very significant $t$-channel-like piece of the reaction
mechanism.  Also in Fig.~\ref{fig:xsec_1405}, in the bin centered at
$W=2.0$~GeV, we show in dashed magenta the flat curve from Williams,
Ji, and Cotanch~\cite{Williams:1991tw}; this was their prediction
based on crossing-symmetry and duality constraints using no
intermediate $N^*$ resonances, and using their lowest estimate for the
$K N \Lambda(1405)$ coupling constant.  Evidently this prediction of
the cross section was too large by at least a factor of two.

Ref~\cite{lineshapepaper} presents the $\Sigma\pi$ mass distributions
(``line shapes'') of the \LambdaOne{} region.  It was not possible to
do a kinematic fit in the $\Sigma^0\pi^0$ analysis to achieve
background rejection as good as for the $\Sigma^\pm\pi^\mp$ channels.
Some $\cos\theta_{K^+}^{c.m.}$ angle-integrated line shape fits had
considerable possibly-incoherent background due to particle
misidentification, higher mass hyperons, or other sources. (See
Figs. 20, 21, and 22 in Ref~\cite{lineshapepaper}.)  For the present
differential cross section analysis, this background could not be
measured separately in each angle bin because of statistical
limitations.  In Ref~\cite{lineshapepaper} the model assumed a linear
background from threshold up to 1.6 GeV, but in fact we do not have
definite knowledge of the background that may have escaped our
simulations near 1.6 GeV, nor its shape under the \LambdaOne.  How
this background is distributed in angle also cannot be determined
accurately from our measurements.  Only an experiment with even better
$\Sigma^0\pi^0$ identification will be able to clarify this point.  In
this article we do not subtract any estimated background from the
\LambdaOne{} cross sections.

\begin{figure}[htpb]
  \includegraphics[width=0.5\textwidth]{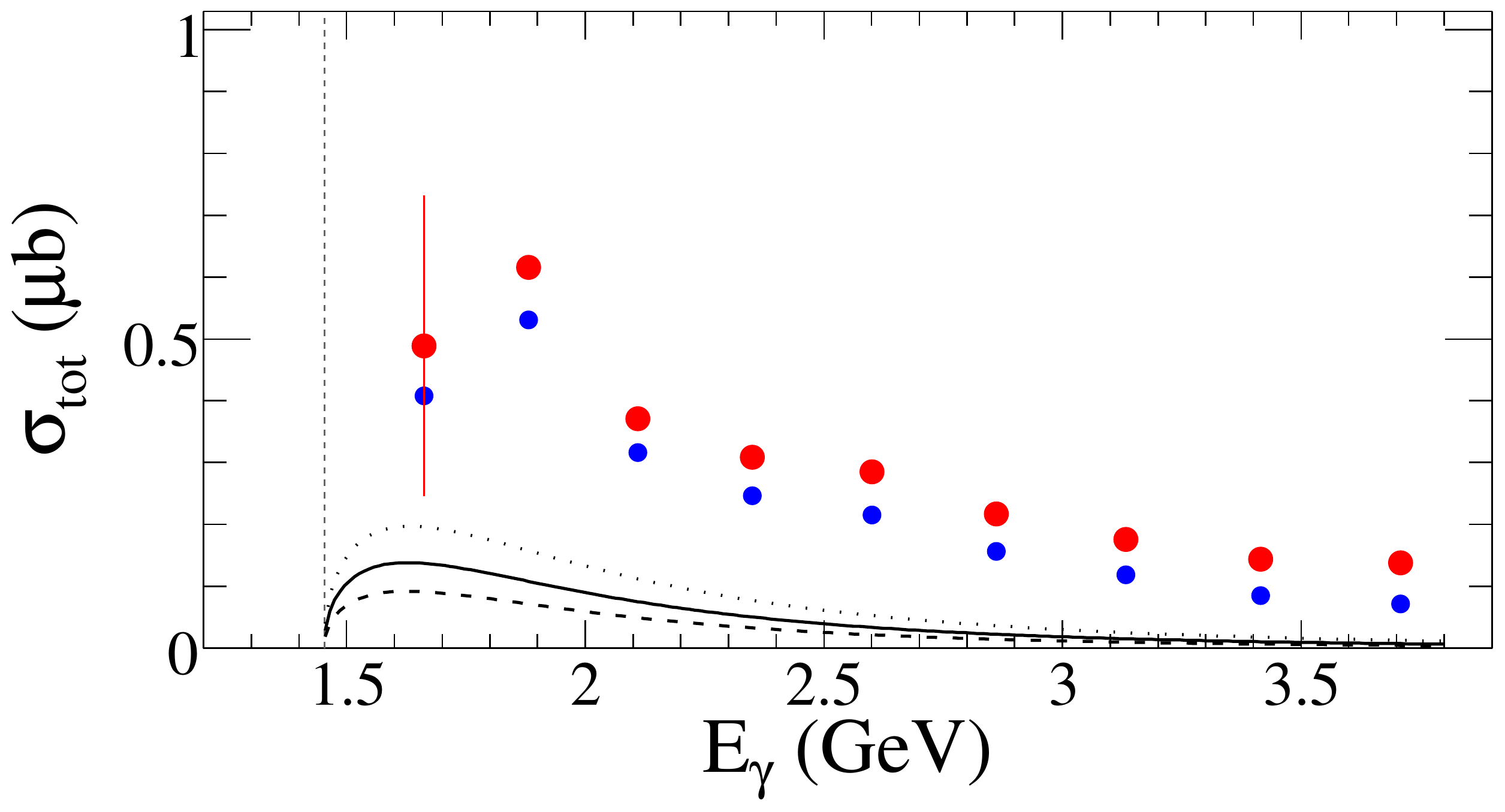}
  \caption{ (Color online) Total cross section of \LambdaOne{} mass
    region versus $E_{\gamma}$. The small blue points show the summed
    measured data from the CLAS detector, while the large red points
    show the extrapolation to all angles. The error bars are combined
    statistical and fitting uncertainties. The vertical dashed line
    shows the reaction threshold. The black curves are a calculation
    by Nam \etal{} based on ~\cite{Nam:2008jy}, with the variable
    $g_{K^{\ast} N \Lambda^{\ast}}$ set to $0$ (solid), $+3.18$
    (dotted), or $-3.18$ (dashed).  }
  \label{fig:sigmatot1405_full}
\end{figure}

Figure~\ref{fig:sigmatot1405_full} shows the total cross section.  The
small black data points show the CLAS data summed over the useful
acceptance of the detector, and the larger solid red points again show
the extrapolated total cross section.  The curves correspond to
calculations of Nam \etal~\cite{Nam_pc}. Clearly, the calculations do
not match the overall scale of the cross sections, nor the dominant
forward peaking behavior.

\subsection{Comparison of results}
\label{subsection:results:resultsCmp}

\begin{figure*}[htpb]
  \includegraphics[width=0.85\textwidth]{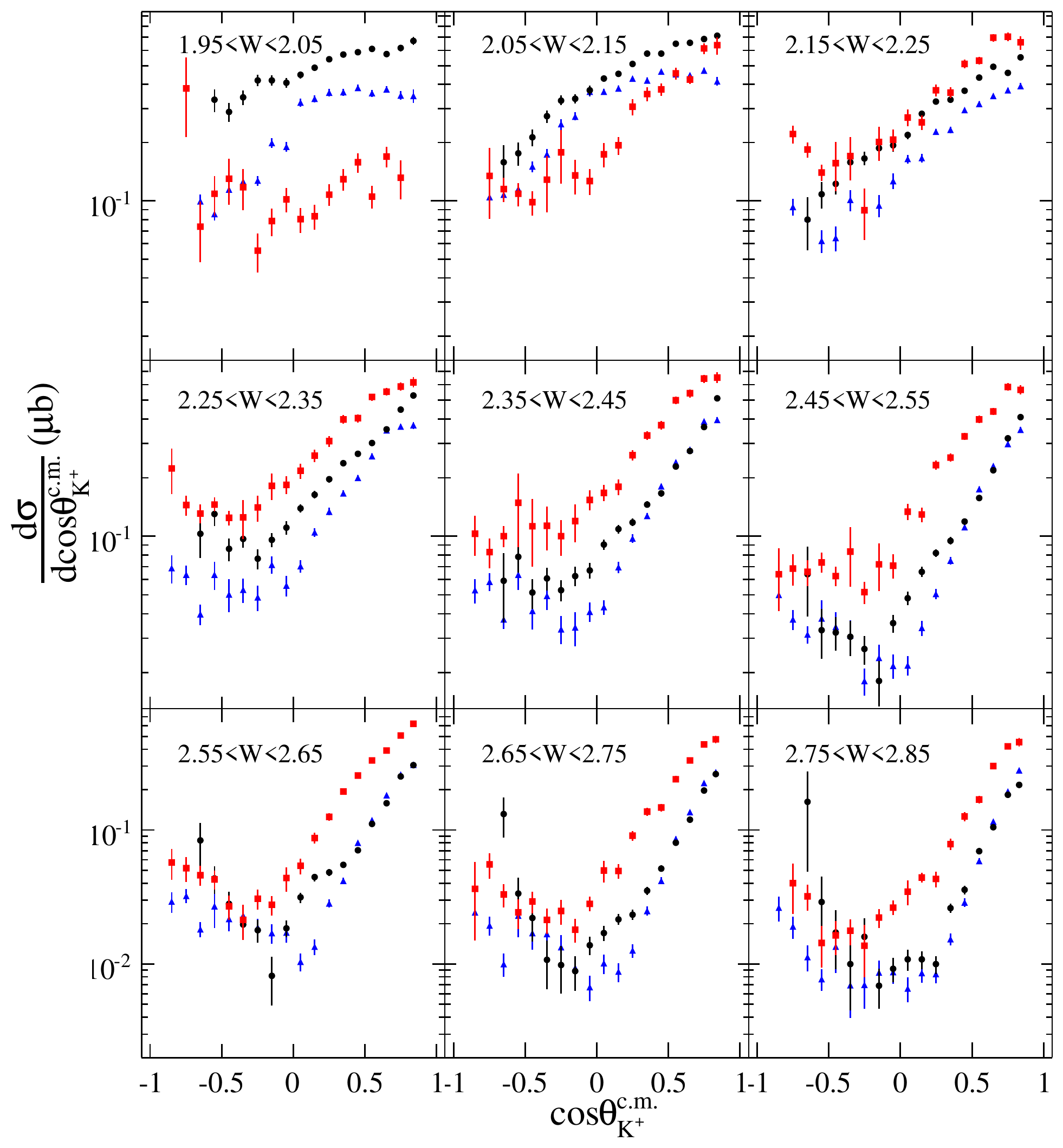}
  \caption{ (Color online) Comparison of the differential cross
    sections of the $\Sigma^{0}(1385)$ (black circles), \LambdaOne{}
    (blue triangles), and \LambdaTwo{} (red squares) as a function of
    kaon production angle $\costhetakp$ for each 100 MeV wide bin in
    energy $W$ (GeV). (For the \LambdaTwo{} the threshold bin has
    $2.02\leq W\leq 2.05$ GeV.)}
  \label{fig:xsec_all}
\end{figure*}

Figure~\ref{fig:xsec_all} shows the differential cross sections for
the $\Sigma^{0}(1385)$, \LambdaTwo, and the \LambdaOne{} region on a
single plot. It is evident that the overall behavior of the three
hyperons is quite similar.  Near threshold the respective phase space
volumes differ and the magnitudes differ greatly. But away from
threshold, each of them falls with increasing kaon production angle
and then rises again in the backward hemisphere. The \LambdaOne{}
region, despite its peculiar charge-dependent line shapes, has a
qualitatively similar differential cross section behavior as the
$\Sigma^{0}(1385)$ and \LambdaTwo.

\begin{figure*}[h!t!b!p]
  \includegraphics[width=\textwidth]{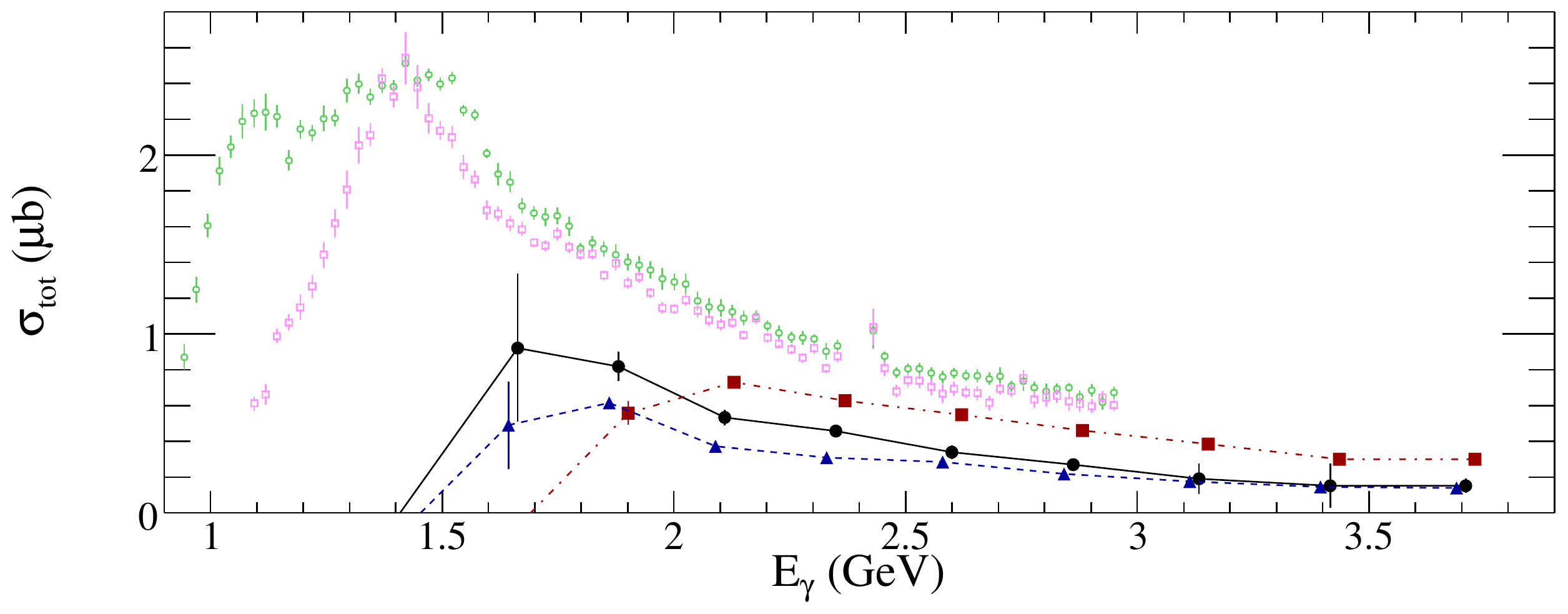}
  \caption{ (Color online) Comparison of total cross sections for the
    $\Sigma^{0}(1385)$ (black circles, solid line), \LambdaOne{} (blue
    triangles, dashed line), and \LambdaTwo{} (red squares, dot-dashed
    line) vs. $E_{\gamma}$. The points for the \LambdaOne{} and
    \LambdaTwo{} have been offset by $20$ MeV for visual clarity. The
    open green circles and open magenta squares are the ground state
    $\Lambda$ and $\Sigma^{0}$ total cross sections, respectively,
    from Ref.~\cite{Bradford_xsec}.}

  \label{fig:sigmatot_full}
\end{figure*}

The total cross sections for all three hyperons are shown together in
Fig.~\ref{fig:sigmatot_full}.  The fall-offs after the initial rise of
the cross sections are similar for the $\Sigma^{0}(1385)$ and the
\LambdaTwo, and slightly slower for the \LambdaOne.  The \LambdaOne{}
is photoproduced with about half the strength of its isospin partner
\LambdaTwo, while the $\Sigma^{0}(1385)$ lies in between.  Recall,
however, that the branching fraction of the \LambdaTwo{} to
$\Sigma\pi$ is 42\%, since it lies above the $N \bar{K}$ threshold,
while the below-threshold \LambdaOne{} decays 100\% to
$\Sigma\pi$. One can therefore say that the \LambdaOne{} and the
\LambdaTwo{} have close to equal strength for decaying to these final
states. We plot for comparison the total photoproduction cross
sections of the ground state $\Lambda$ and
$\SigmaZero$~\cite{Bradford_xsec}. Above about $2.1$~GeV, we see that
cross sections for the excited hyperons differ from those of the
ground state hyperons by factors of less than two to three.

\section{Conclusions}
\label{section:Conclusions}

We have measured for the first time in a single experiment the
differential cross sections for photoproduction of the
$\Sigma^{0}(1385)$, \LambdaTwo, and \LambdaOne.  The
$\Sigma^{0}(1385)$ was measured through its dominant decay mode to
$\Lambda \pizero$, while the other two were measured through all three
$\Sigma\pi$ decay combinations.  They are all of similar size and
angular dependence, and summarized in Figs.~\ref{fig:xsec_all} and
~\ref{fig:sigmatot_full}.  However, the \LambdaOne{} region is
qualitatively different because the angular dependencies of the three
charge combinations are quite different, as shown in
Fig.~\ref{fig:xsec1405_fits}.  This is naturally explained if one
allows strong isospin interference in the production of the
$\Sigma\pi$ final state in the \LambdaOne{} mass region.

This is of particular interest to chiral unitary models that describe
the \LambdaOne{} in terms of a two-pole $I=0$ structure (for a recent
review, see Ref.~\cite{Hyodo:2011ur}).  The strong angular and energy
dependence of the production is a challenge to models that so far use
only a Weinberg-Tomozawa-type contact interaction to generate the
\LambdaOne.  If it is the case that significant $I=1$ strength is
present in the \LambdaOne{} production mechanism, it will be more
difficult to develop a complete theory of its creation, at least in
photoproduction.

Existing model calculations compared to the present results are
moderately successful at reproducing the differential cross sections
of the $\Sigma^{0}(1385)$ and \LambdaTwo.  The models were all
effective Lagrangian calculations with semi-empirical fits of coupling
constants to match the very sparse previous data.  These new data
will, however, allow these and other models to be refined in the
future.  In particular, the role of $K^*$ and $N^*$ contributions to
the reaction mechanisms may now be more closely studied. Much more
work is needed to model the \LambdaOne.

In the future it would be helpful to get much higher statistics for
photoproduction of the \LambdaOne{} to make it possible to do an
isospin analysis in each energy and angle bin separately.  This is
necessary to unravel the nature of this state in more detail.
Nevertheless, the present study has provided the first comprehensive
look at the group of low-lying excited hyperons in photoproduction.
Theoretical understanding of their production and internal structure
should be improved as a result.


\begin{acknowledgments}
We acknowledge the outstanding efforts of the staff of the Accelerator
and Physics Divisions at Jefferson Lab that made this experiment
possible. The work of the Medium Energy Physics group at Carnegie
Mellon University was supported by DOE grant DE-FG02-87ER40315.  The
Southeastern Universities Research Association (SURA) operated the
Thomas Jefferson National Accelerator Facility for the United States
Department of Energy under contract DE-AC05-84ER40150.  Support was
also provided by the National Science Foundation and the United
Kingdom's Science and Technology Facilities Council (STFC).
\end{acknowledgments}
\vfill

\bibliography{xsec}


\appendix*
\section{Numerical Results}

The total cross section results from the present work are given in
Table~\ref{tab:table_total}.  Each row corresponds to a given photon
energy and gives the corresponding result for $\Sigma(1385)$,
\LambdaOne, and \LambdaTwo.  The differential cross section results
from the present work are given Table~\ref{tab:table_diff}.  Each row
corresponds to a given energy bin $W$ and c.m. kaon production angle
$\costhetakp$ for cross sections in the form $d\sigma /
d \costhetakp$.  The quoted uncertainties are the
statistical errors resulting from the yield fitting, acceptance
calculation, photon normalization, and the extrapolation to full
acceptance.  Systematic uncertainties were discussed in the main text.
A zero value for a cross section means no data point was extracted at
that energy.  Electronic tabulations of the results are available from
several archival sources: Refs.~\cite{clasdb}, \cite{durham},
\cite{contact}.

\linespread{1.0}
\begin{table*}
  \caption{ Results of CLAS measurements of $\gamma+p \to K^+ + Y^*$.
    The columns marked \sigmatot{} are the total cross sections for the
    $\Sigma^0(1385)$, \LambdaOne, and \LambdaTwo, respectively, while
    the columns marked ``$\pm$'' are the associated standard
    statistical uncertainties.  The cross section units are in $\mu$b.
    \label{tab:table_total}
  }
  \begin{tabular*}{0.60\textwidth}{@{\extracolsep{\fill}}cccccccc}
    \hline\hline
    Index & $E_{\gamma}$ (GeV) & \multicolumn{2}{c}{$\Sigma^0(1385)$} & \multicolumn{2}{c}{$\Lambda(1405)$} & \multicolumn{2}{c}{$\Lambda(1520)$} \\
          &                    & $\sigma_{tot}$ & $\pm$ & $\sigma_{tot}$ & $\pm$ & $\sigma_{tot}$ & $\pm$ \\ 
    \hline
      1) & 1.662  &  0.921  &  0.414  &  0.489  &  0.244  &  0.000  &  0.000 \\
      2) & 1.881  &  0.818  &  0.083  &  0.615  &  0.008  &  0.557  &  0.065 \\
      3) & 2.110  &  0.532  &  0.043  &  0.371  &  0.005  &  0.729  &  0.029 \\
      4) & 2.350  &  0.457  &  0.026  &  0.309  &  0.011  &  0.625  &  0.015 \\
      5) & 2.600  &  0.339  &  0.039  &  0.285  &  0.008  &  0.549  &  0.014 \\
      6) & 2.862  &  0.268  &  0.027  &  0.216  &  0.008  &  0.460  &  0.015 \\
      7) & 3.133  &  0.189  &  0.085  &  0.176  &  0.014  &  0.383  &  0.017 \\
      8) & 3.416  &  0.151  &  0.124  &  0.144  &  0.014  &  0.299  &  0.016 \\
      9) & 3.709  &  0.151  &  0.041  &  0.138  &  0.010  &  0.297  &  0.032 \\
    \hline\hline
  \end{tabular*}
\end{table*}

\clearpage


\begingroup
\begin{longtable*}{ccccccccccccc}
  \caption{Results of CLAS measurements of $\gamma+p \to K^+ + Y^*$.
    Most energy bins are 100 MeV wide in $W$, centered on the value
    given in the second column.  The threshold $W$ bin for the
    \LambdaTwo{} is only 30 MeV wide.  Most angle bins are 0.1 wide in
    $\costhetakp$, centered on the value given in the third column.
    The columns marked $d\sigma/d\costhetakp$ are the differential
    cross sections for the $\Sigma^0(1385)$, \LambdaTwo, and
    \LambdaOne, respectively, while the columns marked ``$\pm$'' are
    the associated total point-to-point uncertainties.  The cross
    section units are in $\mu$b.  \label{tab:table_diff} }\\

  \hline \hline
  Index & $W$ (GeV) & $\costhetakp$
  & \multicolumn{2}{c}{$\Sigma^0(1385)$}
  & \multicolumn{2}{c}{$\Lambda(1520)$}
  & \multicolumn{2}{c}{$\Lambda(1405) \to \SigmaPlus \pim$}
  & \multicolumn{2}{c}{$\Lambda(1405) \to \SigmaZero \pizero$}
  & \multicolumn{2}{c}{$\Lambda(1405) \to \SigmaMinus \pip$} \\
  & & & $d\sigma/d\costhetakp$ & $\pm$ & $d\sigma/d\costhetakp$ & $\pm$ & $d\sigma/d\costhetakp$ & $\pm$ & $d\sigma/d\costhetakp$ & $\pm$ & $d\sigma/d\costhetakp$ & $\pm$ \\
  \hline  
  \endfirsthead

  \multicolumn{13}{c}
   {{\bfseries \tablename\ \thetable{} -- continued}} \\

  \hline \hline
  Index & $W$ (GeV) & $\costhetakp$
  & \multicolumn{2}{c}{$\Sigma^0(1385)$}
  & \multicolumn{2}{c}{$\Lambda(1520)$}
  & \multicolumn{2}{c}{$\Lambda(1405) \to \SigmaPlus \pim$}
  & \multicolumn{2}{c}{$\Lambda(1405) \to \SigmaZero \pizero$}
  & \multicolumn{2}{c}{$\Lambda(1405) \to \SigmaMinus \pip$} \\
  & & & $d\sigma/d\costhetakp$ & $\pm$ & $d\sigma/d\costhetakp$ & $\pm$ & $d\sigma/d\costhetakp$ & $\pm$ & $d\sigma/d\costhetakp$ & $\pm$ & $d\sigma/d\costhetakp$ & $\pm$ \\
  \hline  
  \endhead

  \hline\hline
  \endlastfoot

   1) & 2.000 & -0.85 & 0.0000 & 0.0000 &          &       & 0.0000 & 0.0000 & 0.0000 & 0.0000 & 0.0000 & 0.0000 \\
   2) & 2.000 & -0.75 & 0.0000 & 0.0000 &          &       & 0.0000 & 0.0000 & 0.0000 & 0.0000 & 0.0000 & 0.0000 \\
   3) & 2.000 & -0.65 & 0.0000 & 0.0000 &          &       & 0.0471 & 0.0062 & 0.0000 & 0.0000 & 0.0523 & 0.0056 \\
   4) & 2.000 & -0.55 & 0.3327 & 0.0440 &          &       & 0.0478 & 0.0053 & 0.0000 & 0.0000 & 0.0379 & 0.0039 \\
   5) & 2.000 & -0.45 & 0.2887 & 0.0324 &          &       & 0.0683 & 0.0058 & 0.0000 & 0.0000 & 0.0462 & 0.0037 \\
   6) & 2.000 & -0.35 & 0.3424 & 0.0299 &          &       & 0.0771 & 0.0059 & 0.0000 & 0.0000 & 0.0486 & 0.0039 \\
   7) & 2.000 & -0.25 & 0.4201 & 0.0290 &          &       & 0.0707 & 0.0054 & 0.0054 & 0.0033 & 0.0509 & 0.0039 \\
   8) & 2.000 & -0.15 & 0.4202 & 0.0258 &          &       & 0.1027 & 0.0067 & 0.0374 & 0.0081 & 0.0596 & 0.0041 \\
   9) & 2.000 & -0.05 & 0.4074 & 0.0240 &          &       & 0.0893 & 0.0058 & 0.0369 & 0.0086 & 0.0642 & 0.0039 \\
  10) & 2.000 &  0.05 & 0.4483 & 0.0188 &          &       & 0.1074 & 0.0048 & 0.1481 & 0.0146 & 0.0668 & 0.0031 \\
  11) & 2.000 &  0.15 & 0.4882 & 0.0175 &          &       & 0.1214 & 0.0056 & 0.1439 & 0.0131 & 0.0712 & 0.0030 \\
  12) & 2.000 &  0.25 & 0.5392 & 0.0175 &          &       & 0.1304 & 0.0054 & 0.1571 & 0.0145 & 0.0752 & 0.0030 \\
  13) & 2.000 &  0.35 & 0.5712 & 0.0182 &          &       & 0.1198 & 0.0054 & 0.1597 & 0.0134 & 0.0844 & 0.0033 \\
  14) & 2.000 &  0.45 & 0.5861 & 0.0176 &          &       & 0.1219 & 0.0047 & 0.1748 & 0.0137 & 0.0860 & 0.0032 \\
  15) & 2.000 &  0.55 & 0.6110 & 0.0181 &          &       & 0.1124 & 0.0049 & 0.1537 & 0.0130 & 0.0936 & 0.0033 \\
  16) & 2.000 &  0.65 & 0.5737 & 0.0187 &          &       & 0.1174 & 0.0044 & 0.1572 & 0.0140 & 0.1032 & 0.0036 \\
  17) & 2.000 &  0.75 & 0.6170 & 0.0223 &          &       & 0.1082 & 0.0043 & 0.1302 & 0.0162 & 0.1121 & 0.0038 \\
  18) & 2.000 &  0.84 & 0.6719 & 0.0342 &          &       & 0.1111 & 0.0077 & 0.0000 & 0.0000 & 0.1073 & 0.0055 \\
& \\[0.1em]
  19) & 2.035 & -0.85 &        &        & 0.0000 & 0.0000 \\
  20) & 2.035 & -0.75 &        &        & 0.3818 & 0.1687 \\
  21) & 2.035 & -0.65 &        &        & 0.0736 & 0.0253 \\
  22) & 2.035 & -0.55 &        &        & 0.1091 & 0.0251 \\
  23) & 2.035 & -0.45 &        &        & 0.1303 & 0.0350 \\
  24) & 2.035 & -0.35 &        &        & 0.1177 & 0.0282 \\
  25) & 2.035 & -0.25 &        &        & 0.0553 & 0.0127 \\
  26) & 2.035 & -0.15 &        &        & 0.0786 & 0.0127 \\
  27) & 2.035 & -0.05 &        &        & 0.1016 & 0.0148 \\
  28) & 2.035 &  0.05 &        &        & 0.0802 & 0.0120 \\
  29) & 2.035 &  0.15 &        &        & 0.0834 & 0.0121 \\
  30) & 2.035 &  0.25 &        &        & 0.1078 & 0.0140 \\
  31) & 2.035 &  0.35 &        &        & 0.1296 & 0.0164 \\
  32) & 2.035 &  0.45 &        &        & 0.1583 & 0.0180 \\
  33) & 2.035 &  0.55 &        &        & 0.1054 & 0.0145 \\
  34) & 2.035 &  0.65 &        &        & 0.1694 & 0.0209 \\
  35) & 2.035 &  0.75 &        &        & 0.1316 & 0.0299 \\
  36) & 2.035 &  0.84 &        &        & 0.0000 & 0.0000 \\
& \\[0.1em]
  37) & 2.100 & -0.85 & 0.0000 & 0.0000 & 0.0000 & 0.0000 & 0.0000 & 0.0000 & 0.0000 & 0.0000 & 0.0000 & 0.0000 \\
  38) & 2.100 & -0.75 & 0.0000 & 0.0000 & 0.1347 & 0.0536 & 0.0927 & 0.0121 & 0.0000 & 0.0000 & 0.0117 & 0.0042 \\
  39) & 2.100 & -0.65 & 0.1585 & 0.0357 & 0.1150 & 0.0168 & 0.0926 & 0.0071 & 0.0000 & 0.0000 & 0.0152 & 0.0028 \\
  40) & 2.100 & -0.55 & 0.1757 & 0.0245 & 0.1086 & 0.0150 & 0.0934 & 0.0069 & 0.0000 & 0.0000 & 0.0219 & 0.0027 \\
  41) & 2.100 & -0.45 & 0.2133 & 0.0214 & 0.0982 & 0.0139 & 0.0993 & 0.0062 & 0.0207 & 0.0070 & 0.0301 & 0.0027 \\
  42) & 2.100 & -0.35 & 0.2733 & 0.0200 & 0.1287 & 0.0418 & 0.0936 & 0.0064 & 0.0460 & 0.0083 & 0.0345 & 0.0028 \\
  43) & 2.100 & -0.25 & 0.3303 & 0.0194 & 0.1782 & 0.0552 & 0.1102 & 0.0063 & 0.0965 & 0.0150 & 0.0425 & 0.0031 \\
  44) & 2.100 & -0.15 & 0.3370 & 0.0185 & 0.1353 & 0.0274 & 0.1249 & 0.0068 & 0.1017 & 0.0107 & 0.0466 & 0.0030 \\
  45) & 2.100 & -0.05 & 0.3729 & 0.0216 & 0.1264 & 0.0193 & 0.1420 & 0.0074 & 0.1679 & 0.0181 & 0.0561 & 0.0035 \\
  46) & 2.100 &  0.05 & 0.4288 & 0.0152 & 0.1742 & 0.0250 & 0.1510 & 0.0058 & 0.1558 & 0.0104 & 0.0597 & 0.0027 \\
  47) & 2.100 &  0.15 & 0.4534 & 0.0147 & 0.1934 & 0.0213 & 0.1637 & 0.0056 & 0.1567 & 0.0106 & 0.0606 & 0.0025 \\
  48) & 2.100 &  0.25 & 0.5092 & 0.0151 & 0.3059 & 0.0286 & 0.1752 & 0.0052 & 0.1832 & 0.0108 & 0.0695 & 0.0026 \\
  49) & 2.100 &  0.35 & 0.5776 & 0.0150 & 0.3563 & 0.0293 & 0.1811 & 0.0057 & 0.1631 & 0.0104 & 0.0760 & 0.0026 \\
  50) & 2.100 &  0.45 & 0.5759 & 0.0146 & 0.3760 & 0.0283 & 0.1843 & 0.0049 & 0.1916 & 0.0096 & 0.0894 & 0.0027 \\
  51) & 2.100 &  0.55 & 0.6492 & 0.0161 & 0.4561 & 0.0313 & 0.1580 & 0.0045 & 0.1900 & 0.0102 & 0.1027 & 0.0029 \\
  52) & 2.100 &  0.65 & 0.6577 & 0.0172 & 0.4245 & 0.0248 & 0.1371 & 0.0046 & 0.1858 & 0.0117 & 0.1231 & 0.0034 \\
  53) & 2.100 &  0.75 & 0.6869 & 0.0198 & 0.6135 & 0.0396 & 0.1277 & 0.0046 & 0.2033 & 0.0143 & 0.1417 & 0.0043 \\
  54) & 2.100 &  0.84 & 0.7114 & 0.0280 & 0.6366 & 0.0652 & 0.0919 & 0.0061 & 0.1778 & 0.0188 & 0.1459 & 0.0049 \\
& \\[0.1em]
  55) & 2.200 & -0.85 & 0.0000 & 0.0000 & 0.0000 & 0.0000 & 0.0000 & 0.0000 & 0.0000 & 0.0000 & 0.0090 & 0.0045 \\
  56) & 2.200 & -0.75 & 0.0000 & 0.0000 & 0.2219 & 0.0236 & 0.0766 & 0.0086 & 0.0000 & 0.0000 & 0.0166 & 0.0031 \\
  57) & 2.200 & -0.65 & 0.0802 & 0.0246 & 0.1840 & 0.0162 & 0.0648 & 0.0066 & 0.0000 & 0.0000 & 0.0164 & 0.0027 \\
  58) & 2.200 & -0.55 & 0.1080 & 0.0169 & 0.1402 & 0.0139 & 0.0426 & 0.0057 & 0.0092 & 0.0056 & 0.0103 & 0.0020 \\
  59) & 2.200 & -0.45 & 0.1225 & 0.0145 & 0.1566 & 0.0447 & 0.0319 & 0.0045 & 0.0187 & 0.0080 & 0.0137 & 0.0021 \\
  60) & 2.200 & -0.35 & 0.1582 & 0.0133 & 0.1707 & 0.0433 & 0.0274 & 0.0047 & 0.0501 & 0.0117 & 0.0237 & 0.0024 \\
  61) & 2.200 & -0.25 & 0.1660 & 0.0128 & 0.0894 & 0.0264 & 0.0195 & 0.0042 & 0.0490 & 0.0097 & 0.0210 & 0.0024 \\
  62) & 2.200 & -0.15 & 0.1873 & 0.0118 & 0.2019 & 0.0410 & 0.0220 & 0.0036 & 0.0488 & 0.0117 & 0.0238 & 0.0024 \\
  63) & 2.200 & -0.05 & 0.1936 & 0.0134 & 0.2077 & 0.0263 & 0.0349 & 0.0046 & 0.0622 & 0.0103 & 0.0294 & 0.0027 \\
  64) & 2.200 &  0.05 & 0.2185 & 0.0097 & 0.2698 & 0.0257 & 0.0456 & 0.0042 & 0.0842 & 0.0079 & 0.0346 & 0.0020 \\
  65) & 2.200 &  0.15 & 0.2817 & 0.0102 & 0.2544 & 0.0227 & 0.0620 & 0.0039 & 0.0710 & 0.0077 & 0.0334 & 0.0020 \\
  66) & 2.200 &  0.25 & 0.3265 & 0.0108 & 0.3722 & 0.0256 & 0.0889 & 0.0047 & 0.0999 & 0.0070 & 0.0386 & 0.0021 \\
  67) & 2.200 &  0.35 & 0.3327 & 0.0102 & 0.3629 & 0.0223 & 0.1037 & 0.0043 & 0.0763 & 0.0067 & 0.0527 & 0.0022 \\
  68) & 2.200 &  0.45 & 0.3703 & 0.0101 & 0.5110 & 0.0256 & 0.1104 & 0.0040 & 0.1168 & 0.0070 & 0.0678 & 0.0024 \\
  69) & 2.200 &  0.55 & 0.4328 & 0.0116 & 0.5309 & 0.0245 & 0.1116 & 0.0046 & 0.1201 & 0.0073 & 0.0846 & 0.0026 \\
  70) & 2.200 &  0.65 & 0.4921 & 0.0131 & 0.6959 & 0.0308 & 0.1133 & 0.0043 & 0.1327 & 0.0079 & 0.1020 & 0.0030 \\
  71) & 2.200 &  0.75 & 0.4580 & 0.0143 & 0.7051 & 0.0350 & 0.1085 & 0.0048 & 0.1531 & 0.0098 & 0.1106 & 0.0034 \\
  72) & 2.200 &  0.84 & 0.5509 & 0.0202 & 0.6578 & 0.0532 & 0.0810 & 0.0049 & 0.1886 & 0.0142 & 0.1234 & 0.0047 \\
& \\[0.1em]
  73) & 2.300 & -0.85 & 0.0000 & 0.0000 & 0.2232 & 0.0584 & 0.0396 & 0.0088 & 0.0000 & 0.0000 & 0.0289 & 0.0070 \\
  74) & 2.300 & -0.75 & 0.0000 & 0.0000 & 0.1442 & 0.0169 & 0.0335 & 0.0066 & 0.0000 & 0.0000 & 0.0299 & 0.0032 \\
  75) & 2.300 & -0.65 & 0.1029 & 0.0254 & 0.1310 & 0.0148 & 0.0203 & 0.0038 & 0.0000 & 0.0000 & 0.0195 & 0.0027 \\
  76) & 2.300 & -0.55 & 0.1300 & 0.0173 & 0.1458 & 0.0123 & 0.0267 & 0.0032 & 0.0244 & 0.0097 & 0.0124 & 0.0020 \\
  77) & 2.300 & -0.45 & 0.0861 & 0.0114 & 0.1244 & 0.0112 & 0.0187 & 0.0026 & 0.0167 & 0.0090 & 0.0150 & 0.0018 \\
  78) & 2.300 & -0.35 & 0.0971 & 0.0099 & 0.1252 & 0.0281 & 0.0175 & 0.0026 & 0.0170 & 0.0070 & 0.0187 & 0.0021 \\
  79) & 2.300 & -0.25 & 0.0768 & 0.0091 & 0.1405 & 0.0212 & 0.0086 & 0.0019 & 0.0188 & 0.0069 & 0.0212 & 0.0020 \\
  80) & 2.300 & -0.15 & 0.0959 & 0.0081 & 0.1818 & 0.0283 & 0.0264 & 0.0026 & 0.0245 & 0.0065 & 0.0206 & 0.0020 \\
  81) & 2.300 & -0.05 & 0.1108 & 0.0087 & 0.1836 & 0.0192 & 0.0129 & 0.0021 & 0.0229 & 0.0060 & 0.0201 & 0.0019 \\
  82) & 2.300 &  0.05 & 0.1394 & 0.0071 & 0.2169 & 0.0195 & 0.0273 & 0.0021 & 0.0215 & 0.0044 & 0.0214 & 0.0016 \\
  83) & 2.300 &  0.15 & 0.1632 & 0.0072 & 0.2596 & 0.0197 & 0.0424 & 0.0024 & 0.0351 & 0.0047 & 0.0274 & 0.0017 \\
  84) & 2.300 &  0.25 & 0.1958 & 0.0074 & 0.3080 & 0.0198 & 0.0507 & 0.0026 & 0.0438 & 0.0051 & 0.0396 & 0.0019 \\
  85) & 2.300 &  0.35 & 0.2373 & 0.0073 & 0.3974 & 0.0197 & 0.0632 & 0.0026 & 0.0519 & 0.0048 & 0.0513 & 0.0019 \\
  86) & 2.300 &  0.45 & 0.2652 & 0.0078 & 0.4039 & 0.0194 & 0.0699 & 0.0039 & 0.0592 & 0.0050 & 0.0708 & 0.0022 \\
  87) & 2.300 &  0.55 & 0.3007 & 0.0086 & 0.5180 & 0.0229 & 0.0841 & 0.0039 & 0.0814 & 0.0062 & 0.0927 & 0.0025 \\
  88) & 2.300 &  0.65 & 0.3534 & 0.0103 & 0.5526 & 0.0240 & 0.0953 & 0.0051 & 0.1353 & 0.0073 & 0.1184 & 0.0030 \\
  89) & 2.300 &  0.75 & 0.4469 & 0.0123 & 0.5873 & 0.0288 & 0.0779 & 0.0050 & 0.1637 & 0.0081 & 0.1241 & 0.0032 \\
  90) & 2.300 &  0.84 & 0.5282 & 0.0174 & 0.6168 & 0.0374 & 0.0723 & 0.0069 & 0.1685 & 0.0123 & 0.1306 & 0.0043 \\
& \\[0.1em]
  91) & 2.400 & -0.85 & 0.0000 & 0.0000 & 0.1033 & 0.0241 & 0.0179 & 0.0055 & 0.0000 & 0.0000 & 0.0350 & 0.0054 \\
  92) & 2.400 & -0.75 & 0.0000 & 0.0000 & 0.0830 & 0.0144 & 0.0222 & 0.0053 & 0.0000 & 0.0000 & 0.0362 & 0.0033 \\
  93) & 2.400 & -0.65 & 0.0591 & 0.0231 & 0.1001 & 0.0122 & 0.0092 & 0.0028 & 0.0000 & 0.0000 & 0.0282 & 0.0028 \\
  94) & 2.400 & -0.55 & 0.0785 & 0.0135 & 0.1486 & 0.0610 & 0.0185 & 0.0033 & 0.0255 & 0.0098 & 0.0194 & 0.0021 \\
  95) & 2.400 & -0.45 & 0.0514 & 0.0091 & 0.1126 & 0.0429 & 0.0075 & 0.0018 & 0.0200 & 0.0079 & 0.0140 & 0.0017 \\
  96) & 2.400 & -0.35 & 0.0609 & 0.0083 & 0.1135 & 0.0288 & 0.0121 & 0.0019 & 0.0191 & 0.0070 & 0.0182 & 0.0018 \\
  97) & 2.400 & -0.25 & 0.0530 & 0.0066 & 0.1002 & 0.0208 & 0.0073 & 0.0016 & 0.0127 & 0.0050 & 0.0134 & 0.0017 \\
  98) & 2.400 & -0.15 & 0.0625 & 0.0071 & 0.1194 & 0.0263 & 0.0117 & 0.0019 & 0.0168 & 0.0063 & 0.0055 & 0.0016 \\
  99) & 2.400 & -0.05 & 0.0669 & 0.0059 & 0.1538 & 0.0177 & 0.0118 & 0.0018 & 0.0178 & 0.0042 & 0.0115 & 0.0013 \\
 100) & 2.400 &  0.05 & 0.0907 & 0.0054 & 0.1674 & 0.0161 & 0.0148 & 0.0015 & 0.0153 & 0.0032 & 0.0134 & 0.0014 \\
 101) & 2.400 &  0.15 & 0.1088 & 0.0059 & 0.1796 & 0.0170 & 0.0296 & 0.0018 & 0.0221 & 0.0040 & 0.0176 & 0.0015 \\
 102) & 2.400 &  0.25 & 0.1180 & 0.0055 & 0.2611 & 0.0164 & 0.0327 & 0.0020 & 0.0353 & 0.0040 & 0.0296 & 0.0016 \\
 103) & 2.400 &  0.35 & 0.1453 & 0.0055 & 0.3293 & 0.0174 & 0.0414 & 0.0020 & 0.0446 & 0.0040 & 0.0409 & 0.0018 \\
 104) & 2.400 &  0.45 & 0.1661 & 0.0061 & 0.3724 & 0.0178 & 0.0642 & 0.0023 & 0.0533 & 0.0047 & 0.0627 & 0.0020 \\
 105) & 2.400 &  0.55 & 0.2275 & 0.0073 & 0.4989 & 0.0225 & 0.0726 & 0.0030 & 0.0815 & 0.0052 & 0.0855 & 0.0024 \\
 106) & 2.400 &  0.65 & 0.2746 & 0.0090 & 0.5422 & 0.0256 & 0.0905 & 0.0036 & 0.0822 & 0.0064 & 0.1065 & 0.0029 \\
 107) & 2.400 &  0.75 & 0.3637 & 0.0108 & 0.6443 & 0.0300 & 0.1096 & 0.0043 & 0.1486 & 0.0075 & 0.1301 & 0.0033 \\
 108) & 2.400 &  0.84 & 0.5110 & 0.0160 & 0.6536 & 0.0389 & 0.0994 & 0.0062 & 0.1660 & 0.0106 & 0.1305 & 0.0043 \\
& \\[0.1em]
 109) & 2.500 & -0.85 & 0.0000 & 0.0000 & 0.0639 & 0.0227 & 0.0197 & 0.0054 & 0.0000 & 0.0000 & 0.0303 & 0.0046 \\
 110) & 2.500 & -0.75 & 0.0000 & 0.0000 & 0.0685 & 0.0126 & 0.0152 & 0.0035 & 0.0000 & 0.0000 & 0.0222 & 0.0028 \\
 111) & 2.500 & -0.65 & 0.0638 & 0.0250 & 0.0656 & 0.0099 & 0.0121 & 0.0024 & 0.0000 & 0.0000 & 0.0193 & 0.0022 \\
 112) & 2.500 & -0.55 & 0.0329 & 0.0094 & 0.0736 & 0.0088 & 0.0127 & 0.0023 & 0.0106 & 0.0087 & 0.0146 & 0.0017 \\
 113) & 2.500 & -0.45 & 0.0322 & 0.0064 & 0.0625 & 0.0072 & 0.0083 & 0.0013 & 0.0156 & 0.0061 & 0.0103 & 0.0014 \\
 114) & 2.500 & -0.35 & 0.0305 & 0.0063 & 0.0833 & 0.0282 & 0.0131 & 0.0018 & 0.0092 & 0.0055 & 0.0087 & 0.0014 \\
 115) & 2.500 & -0.25 & 0.0264 & 0.0045 & 0.0516 & 0.0068 & 0.0067 & 0.0011 & 0.0048 & 0.0024 & 0.0067 & 0.0012 \\
 116) & 2.500 & -0.15 & 0.0181 & 0.0047 & 0.0719 & 0.0196 & 0.0059 & 0.0012 & 0.0082 & 0.0035 & 0.0097 & 0.0014 \\
 117) & 2.500 & -0.05 & 0.0357 & 0.0038 & 0.0707 & 0.0101 & 0.0061 & 0.0010 & 0.0100 & 0.0028 & 0.0054 & 0.0009 \\
 118) & 2.500 &  0.05 & 0.0481 & 0.0038 & 0.1340 & 0.0132 & 0.0082 & 0.0010 & 0.0085 & 0.0021 & 0.0050 & 0.0009 \\
 119) & 2.500 &  0.15 & 0.0658 & 0.0041 & 0.1291 & 0.0108 & 0.0102 & 0.0010 & 0.0151 & 0.0026 & 0.0087 & 0.0010 \\
 120) & 2.500 &  0.25 & 0.0821 & 0.0039 & 0.2324 & 0.0130 & 0.0176 & 0.0011 & 0.0199 & 0.0027 & 0.0132 & 0.0011 \\
 121) & 2.500 &  0.35 & 0.0949 & 0.0042 & 0.2539 & 0.0124 & 0.0230 & 0.0013 & 0.0277 & 0.0029 & 0.0244 & 0.0013 \\
 122) & 2.500 &  0.45 & 0.1190 & 0.0046 & 0.3256 & 0.0145 & 0.0392 & 0.0016 & 0.0303 & 0.0031 & 0.0421 & 0.0016 \\
 123) & 2.500 &  0.55 & 0.1570 & 0.0057 & 0.3979 & 0.0177 & 0.0529 & 0.0021 & 0.0592 & 0.0041 & 0.0622 & 0.0020 \\
 124) & 2.500 &  0.65 & 0.2177 & 0.0071 & 0.4366 & 0.0193 & 0.0765 & 0.0025 & 0.0719 & 0.0048 & 0.0808 & 0.0024 \\
 125) & 2.500 &  0.75 & 0.3179 & 0.0092 & 0.5848 & 0.0255 & 0.0913 & 0.0030 & 0.1058 & 0.0060 & 0.1009 & 0.0028 \\
 126) & 2.500 &  0.84 & 0.4095 & 0.0137 & 0.5648 & 0.0314 & 0.1048 & 0.0049 & 0.1361 & 0.0096 & 0.1110 & 0.0039 \\
& \\[0.1em]
 127) & 2.600 & -0.85 & 0.0000 & 0.0000 & 0.0573 & 0.0150 & 0.0060 & 0.0028 & 0.0000 & 0.0000 & 0.0232 & 0.0044 \\
 128) & 2.600 & -0.75 & 0.0000 & 0.0000 & 0.0521 & 0.0109 & 0.0129 & 0.0030 & 0.0000 & 0.0000 & 0.0194 & 0.0023 \\
 129) & 2.600 & -0.65 & 0.0832 & 0.0299 & 0.0461 & 0.0078 & 0.0068 & 0.0017 & 0.0000 & 0.0000 & 0.0113 & 0.0017 \\
 130) & 2.600 & -0.55 & 0.0435 & 0.0102 & 0.0428 & 0.0069 & 0.0100 & 0.0017 & 0.0112 & 0.0079 & 0.0056 & 0.0012 \\
 131) & 2.600 & -0.45 & 0.0280 & 0.0065 & 0.0270 & 0.0056 & 0.0090 & 0.0013 & 0.0043 & 0.0038 & 0.0084 & 0.0011 \\
 132) & 2.600 & -0.35 & 0.0197 & 0.0045 & 0.0214 & 0.0063 & 0.0083 & 0.0013 & 0.0075 & 0.0029 & 0.0072 & 0.0011 \\
 133) & 2.600 & -0.25 & 0.0178 & 0.0035 & 0.0306 & 0.0051 & 0.0080 & 0.0010 & 0.0056 & 0.0026 & 0.0052 & 0.0010 \\
 134) & 2.600 & -0.15 & 0.0081 & 0.0032 & 0.0276 & 0.0047 & 0.0046 & 0.0008 & 0.0046 & 0.0025 & 0.0079 & 0.0010 \\
 135) & 2.600 & -0.05 & 0.0185 & 0.0027 & 0.0437 & 0.0090 & 0.0044 & 0.0006 & 0.0114 & 0.0026 & 0.0013 & 0.0006 \\
 136) & 2.600 &  0.05 & 0.0314 & 0.0028 & 0.0539 & 0.0070 & 0.0036 & 0.0007 & 0.0041 & 0.0014 & 0.0027 & 0.0006 \\
 137) & 2.600 &  0.15 & 0.0444 & 0.0028 & 0.0867 & 0.0080 & 0.0055 & 0.0007 & 0.0058 & 0.0016 & 0.0022 & 0.0006 \\
 138) & 2.600 &  0.25 & 0.0481 & 0.0028 & 0.1252 & 0.0080 & 0.0099 & 0.0008 & 0.0125 & 0.0017 & 0.0060 & 0.0007 \\
 139) & 2.600 &  0.35 & 0.0548 & 0.0030 & 0.1944 & 0.0100 & 0.0112 & 0.0008 & 0.0181 & 0.0021 & 0.0125 & 0.0009 \\
 140) & 2.600 &  0.45 & 0.0706 & 0.0034 & 0.2559 & 0.0115 & 0.0255 & 0.0012 & 0.0298 & 0.0024 & 0.0247 & 0.0012 \\
 141) & 2.600 &  0.55 & 0.1106 & 0.0046 & 0.3306 & 0.0159 & 0.0427 & 0.0016 & 0.0340 & 0.0031 & 0.0412 & 0.0016 \\
 142) & 2.600 &  0.65 & 0.1580 & 0.0055 & 0.3906 & 0.0178 & 0.0649 & 0.0020 & 0.0555 & 0.0039 & 0.0606 & 0.0019 \\
 143) & 2.600 &  0.75 & 0.2498 & 0.0078 & 0.5053 & 0.0214 & 0.0922 & 0.0029 & 0.0903 & 0.0052 & 0.0773 & 0.0024 \\
 144) & 2.600 &  0.84 & 0.3048 & 0.0118 & 0.6184 & 0.0341 & 0.0982 & 0.0040 & 0.1164 & 0.0081 & 0.0924 & 0.0035 \\
& \\[0.1em]
 145) & 2.700 & -0.85 & 0.0000 & 0.0000 & 0.0362 & 0.0213 & 0.0093 & 0.0043 & 0.0000 & 0.0000 & 0.0151 & 0.0041 \\
 146) & 2.700 & -0.75 & 0.0000 & 0.0000 & 0.0551 & 0.0120 & 0.0074 & 0.0024 & 0.0000 & 0.0000 & 0.0120 & 0.0021 \\
 147) & 2.700 & -0.65 & 0.1310 & 0.0433 & 0.0329 & 0.0066 & 0.0059 & 0.0016 & 0.0000 & 0.0000 & 0.0041 & 0.0012 \\
 148) & 2.700 & -0.55 & 0.0336 & 0.0106 & 0.0243 & 0.0061 & 0.0066 & 0.0014 & 0.0117 & 0.0068 & 0.0045 & 0.0011 \\
 149) & 2.700 & -0.45 & 0.0220 & 0.0073 & 0.0293 & 0.0054 & 0.0063 & 0.0010 & 0.0061 & 0.0040 & 0.0046 & 0.0009 \\
 150) & 2.700 & -0.35 & 0.0107 & 0.0042 & 0.0214 & 0.0046 & 0.0068 & 0.0010 & 0.0057 & 0.0038 & 0.0041 & 0.0008 \\
 151) & 2.700 & -0.25 & 0.0098 & 0.0038 & 0.0248 & 0.0054 & 0.0070 & 0.0014 & 0.0027 & 0.0026 & 0.0037 & 0.0008 \\
 152) & 2.700 & -0.15 & 0.0089 & 0.0026 & 0.0181 & 0.0036 & 0.0030 & 0.0006 & 0.0032 & 0.0018 & 0.0032 & 0.0006 \\
 153) & 2.700 & -0.05 & 0.0138 & 0.0022 & 0.0281 & 0.0035 & 0.0030 & 0.0005 & 0.0015 & 0.0013 & 0.0023 & 0.0005 \\
 154) & 2.700 &  0.05 & 0.0170 & 0.0022 & 0.0498 & 0.0093 & 0.0043 & 0.0006 & 0.0040 & 0.0015 & 0.0018 & 0.0005 \\
 155) & 2.700 &  0.15 & 0.0216 & 0.0021 & 0.0496 & 0.0061 & 0.0038 & 0.0006 & 0.0036 & 0.0012 & 0.0013 & 0.0004 \\
 156) & 2.700 &  0.25 & 0.0233 & 0.0021 & 0.0908 & 0.0078 & 0.0056 & 0.0006 & 0.0042 & 0.0012 & 0.0027 & 0.0005 \\
 157) & 2.700 &  0.35 & 0.0352 & 0.0024 & 0.1365 & 0.0092 & 0.0057 & 0.0007 & 0.0124 & 0.0017 & 0.0068 & 0.0007 \\
 158) & 2.700 &  0.45 & 0.0516 & 0.0030 & 0.1467 & 0.0098 & 0.0147 & 0.0010 & 0.0154 & 0.0021 & 0.0118 & 0.0010 \\
 159) & 2.700 &  0.55 & 0.0802 & 0.0040 & 0.2389 & 0.0135 & 0.0287 & 0.0014 & 0.0310 & 0.0027 & 0.0258 & 0.0014 \\
 160) & 2.700 &  0.65 & 0.1194 & 0.0050 & 0.3309 & 0.0168 & 0.0480 & 0.0018 & 0.0466 & 0.0033 & 0.0413 & 0.0017 \\
 161) & 2.700 &  0.75 & 0.1963 & 0.0072 & 0.4330 & 0.0202 & 0.0720 & 0.0026 & 0.0862 & 0.0051 & 0.0661 & 0.0024 \\
 162) & 2.700 &  0.83 & 0.2617 & 0.0124 & 0.4765 & 0.0317 & 0.0851 & 0.0048 & 0.1070 & 0.0083 & 0.0785 & 0.0040 \\
& \\[0.1em]
 163) & 2.800 & -0.85 & 0.0000 & 0.0000 & 0.0000 & 0.0000 & 0.0036 & 0.0020 & 0.0000 & 0.0000 & 0.0227 & 0.0051 \\
 164) & 2.800 & -0.75 & 0.0000 & 0.0000 & 0.0400 & 0.0163 & 0.0113 & 0.0028 & 0.0000 & 0.0000 & 0.0077 & 0.0022 \\
 165) & 2.800 & -0.65 & 0.1617 & 0.1126 & 0.0320 & 0.0071 & 0.0061 & 0.0021 & 0.0000 & 0.0000 & 0.0052 & 0.0014 \\
 166) & 2.800 & -0.55 & 0.0289 & 0.0160 & 0.0143 & 0.0049 & 0.0052 & 0.0011 & 0.0000 & 0.0000 & 0.0025 & 0.0008 \\
 167) & 2.800 & -0.45 & 0.0172 & 0.0081 & 0.0163 & 0.0046 & 0.0072 & 0.0013 & 0.0044 & 0.0046 & 0.0018 & 0.0006 \\
 168) & 2.800 & -0.35 & 0.0100 & 0.0055 & 0.0177 & 0.0039 & 0.0042 & 0.0009 & 0.0009 & 0.0028 & 0.0019 & 0.0006 \\
 169) & 2.800 & -0.25 & 0.0160 & 0.0059 & 0.0137 & 0.0058 & 0.0035 & 0.0008 & 0.0013 & 0.0021 & 0.0022 & 0.0007 \\
 170) & 2.800 & -0.15 & 0.0069 & 0.0023 & 0.0221 & 0.0036 & 0.0023 & 0.0005 & 0.0034 & 0.0018 & 0.0029 & 0.0005 \\
 171) & 2.800 & -0.05 & 0.0092 & 0.0019 & 0.0264 & 0.0033 & 0.0026 & 0.0006 & 0.0040 & 0.0014 & 0.0021 & 0.0004 \\
 172) & 2.800 &  0.05 & 0.0108 & 0.0019 & 0.0348 & 0.0072 & 0.0019 & 0.0004 & 0.0033 & 0.0013 & 0.0014 & 0.0004 \\
 173) & 2.800 &  0.15 & 0.0108 & 0.0017 & 0.0441 & 0.0036 & 0.0032 & 0.0004 & 0.0035 & 0.0012 & 0.0019 & 0.0004 \\
 174) & 2.800 &  0.25 & 0.0100 & 0.0015 & 0.0429 & 0.0057 & 0.0042 & 0.0005 & 0.0026 & 0.0011 & 0.0016 & 0.0004 \\
 175) & 2.800 &  0.35 & 0.0261 & 0.0021 & 0.0786 & 0.0076 & 0.0044 & 0.0006 & 0.0075 & 0.0013 & 0.0034 & 0.0006 \\
 176) & 2.800 &  0.45 & 0.0357 & 0.0028 & 0.1259 & 0.0098 & 0.0092 & 0.0008 & 0.0104 & 0.0019 & 0.0093 & 0.0009 \\
 177) & 2.800 &  0.55 & 0.0694 & 0.0037 & 0.1690 & 0.0113 & 0.0190 & 0.0012 & 0.0195 & 0.0025 & 0.0201 & 0.0013 \\
 178) & 2.800 &  0.65 & 0.1050 & 0.0048 & 0.2986 & 0.0153 & 0.0376 & 0.0017 & 0.0409 & 0.0033 & 0.0367 & 0.0018 \\
 179) & 2.800 &  0.75 & 0.1832 & 0.0074 & 0.4212 & 0.0209 & 0.0724 & 0.0028 & 0.0645 & 0.0048 & 0.0574 & 0.0026 \\
 180) & 2.800 &  0.83 & 0.2172 & 0.0118 & 0.4534 & 0.0323 & 0.0862 & 0.0050 & 0.1155 & 0.0088 & 0.0770 & 0.0045 \\
  \end{longtable*}
  \endgroup


\end{document}